%% file: main.tex
\documentclass[sigconf, nonacm]{acmart}

\input{header}
\usepackage{pbalance}

\begin{document}


\title{{\toolname}: A Modular Benchmarking Framework for LLM-Enabled NL2SQL Solutions}







\author{Shizheng Hou$^{1}$, Wenqi Pei$^{1}$, Nuo Chen$^{1}$, Quang-Trung Ta$^{1}$,
        Peng Lu$^{2*}$, Beng Chin Ooi$^{2}$}
\affiliation{%
  \institution{$^1$National University of Singapore \quad $^2$Zhejiang University}
}
\email{{shizheng\_hou, wenqi\_pei}@u.nus.edu, nuochen@comp.nus.edu.sg, taqt@nus.edu.sg, {peng.lu, ooibc}@zju.edu.cn}
\thanks{$^*$Corresponding author.}

\input{sections/abstract}

\maketitle

\input{sections/introduction}

\input{sections/background}

\input{sections/methodology}

\input{sections/experiment}

\input{sections/related-work}

\input{sections/discussion}

\input{sections/conclusion}
\onecolumn
\begin{multicols}{2}[]
\bibliographystyle{ACM-Reference-Format}
\bibliography{reference}
\end{multicols}
\twocolumn
\appendix
\input{sections/appendix}

\end{document}

%% file: header.tex
\usepackage{latexsym}

\usepackage[T1]{fontenc}

\usepackage[utf8]{inputenc}

\usepackage{microtype}

\usepackage{balance}
\usepackage{graphicx}

\usepackage{fancyhdr}
\usepackage{enumitem}
\usepackage{utfsym}
\usepackage{amsmath}
\usepackage{pifont}
\usepackage{booktabs}
\usepackage{changepage}

\usepackage{tikz}
\usepackage{filecontents}
\usepackage{comment}
\usepackage{epsfig}
\usepackage{epstopdf}
\usepackage{graphics}
\usepackage{bookmark}
\usepackage{MnSymbol}%

\usepackage{multirow}

\usepackage[ruled, boxed,vlined,linesnumbered]{algorithm2e}
\usepackage{xspace}
\usepackage{subcaption}
\usepackage{verbatim}
\usepackage{mathtools}
\usepackage{upgreek}
\usepackage{cleveref}
\usepackage{multicol,xparse,environ}
\usepackage{caption}
\usepackage{natbib}
\usepackage{amsfonts}
\usepackage[nodisplayskipstretch]{setspace}
\usepackage{textcomp}
\usepackage{bm}
\usepackage{algpseudocode}

\usepackage[most]{tcolorbox}

\newtcolorbox{question}[1][]{%
    colback=black!10,
    colframe=black!10,
    notitle,
    sharp corners,
    borderline west={2pt}{0pt}{red!80!black},
    enhanced,
    breakable,
    before upper=\setlength{\parskip}{0.5em},
    }

\usepackage{lipsum}

\usepackage{makecell}

\usepackage{minted}
\usepackage{float}
\usepackage{listings}
\lstdefinelanguage{json}{
  basicstyle=\ttfamily\small,
  escapeinside={(*@}{@*)},
  breaklines=true,
  captionpos=b,
  frame=single,
  literate=
     *{0}{{{\color{blue}0}}}{1}
      {1}{{{\color{blue}1}}}{1}
      {2}{{{\color{blue}2}}}{1}
      {3}{{{\color{blue}3}}}{1}
      {4}{{{\color{blue}4}}}{1}
      {5}{{{\color{blue}5}}}{1}
      {6}{{{\color{blue}6}}}{1}
      {7}{{{\color{blue}7}}}{1}
      {8}{{{\color{blue}8}}}{1}
      {9}{{{\color{blue}9}}}{1}
      {:}{{{\color{black}{:}}}}{1}
      {,}{{{\color{black}{,}}}}{1}
      {\{}{{{\color{black}{\{}}}}{1}
      {\}}{{{\color{black}{\}}}}}{1}
      {[}{{{\color{black}{[}}}}{1}
      {]}{{{\color{black}{]}}}}{1}
}
\usepackage{xcolor}
\definecolor{backcolour}{rgb}{0.95,0.95,0.92}

\usepackage{hyperref}
\hypersetup{
    colorlinks=true,
    linkcolor=blue,
    urlcolor=blue
}

\usepackage[normalem]{ulem}

\newcommand{\Yes}{\ding{51}}
\newcommand{\No}{\ding{55}}

\definecolor{sota}{RGB}{191, 0, 64}
\definecolor{second}{RGB}{191, 129, 64}

\newcommand{\fst}[1]{\underline{\textbf{#1}}}
\newcommand{\snd}[1]{\underline{#1}}

\def\toolname{\textbf{\texttt{NL2SQLBench}}}

\newtcolorbox{dialogbox}{
  colback=gray!10!white, colframe=black, sharp corners,
  boxrule=0.5mm, top=10pt, bottom=10pt, left=10pt, right=10pt, breakable,
  fontupper=\small
}


%% file: sections/abstract.tex
\begin{abstract}
Natural Language to SQL (NL2SQL) technology empowers non-expert users to query relational databases without requiring SQL expertise.
While large language models (LLMs) have greatly improved NL2SQL algorithms, their rapid development outpaces systematic evaluation, leaving a critical gap in understanding their effectiveness, efficiency, and limitations.
To this end, we present {\toolname}, the first modular evaluation and benchmarking framework for LLM-enabled NL2SQL approaches.
Specifically, we dissect NL2SQL systems into three core modules: Schema Selection, Candidate Generation, and Query Revision.
For each module, we comprehensively review existing strategies and propose novel fine-grained metrics that systematically quantify module-level effectiveness and efficiency.
We further implement these metrics in a flexible multi-agent framework, allowing configurable benchmarking across diverse NL2SQL approaches.
Leveraging {\toolname}, we rigorously evaluate ten representative open-source methods on two datasets, the BIRD development set and the ScienceBenchmark development set, using two LLMs, DeepSeek-V3 and GPT-4o mini.
We systematically assess each approach across the three core modules and evaluate multiple critical performance dimensions.
Our evaluation reveals significant gaps in existing NL2SQL methods, highlighting not only substantial room for accuracy improvements but also the significant computational inefficiency, which severely hampers real-world adoption.
Furthermore, our analysis identifies critical shortcomings in current benchmark datasets and evaluation rules, emphasizing issues such as inaccurate gold SQL annotations and limitations in existing evaluation rules.
By synthesizing these detailed insights into a unified, transparent, and reproducible benchmarking, our study not only establishes a clear reference point for fair comparison across approaches but also serves as essential guidance for future targeted innovation in NL2SQL technology, thus advancing the practical deployment and real-world applicability of NL2SQL technologies.
The {\toolname} project is open-sourced:~\url{https://github.com/neurdb/NL2SQLBench}.
\end{abstract}


%% file: sections/introduction.tex
\section{Introduction}

Natural Language to SQL (NL2SQL) is a pivotal technology that empowers non-expert users to query relational databases using natural language (NL) questions, without requiring any SQL expertise. This technology holds significant value as it democratizes access to data, empowering users to engage with complex database systems and supporting diverse applications across various business domains~\cite{nlq-of-biq,song2024enhancing,demo-dbgpt,llm-from-data,lian2024chatbi,weng2024datalab, maamari2024end}. Concurrently, as noted by~\citet{DBLP:journals/tkde/LiuSLMJZFLTL25}, the provision of NL2SQL solutions has shifted from a conceptual idea to an essential strategy among database vendors. Consequently, the question of how to effectively implement such solutions has become an actively discussed topic~\cite{FloratouPZDHTCA24, ailamaki2025cambridge}.

Recent advancements in Large Language Models (LLMs)~\cite{zhao2023survey, minaee2024largelanguagemodelssurvey, kaddour2023challenges} have introduced a transformative paradigm for NL2SQL tasks, offering new opportunities for enhanced performance and adaptability. Numerous LLM-based approaches~\cite{pourreza2024din, gao2023text, dong2023c3, macsql2025, lee2024mcs, li2024codes, combine-slm-llm, pourreza2024dts, qu2024before, li2024pet, ren2024purple, askari2024magic, maamari2024death, caferouglu2024esql, xiyansql, cogsql, long2025bridging, aid-sql, lyu2025sqlo1, sphinteract, pei2025optimizing} have been proposed, establishing themselves as the most prominent solutions in the NL2SQL landscape.

\begin{figure}[t]
  \centering
  \includegraphics[width=1.0\linewidth]{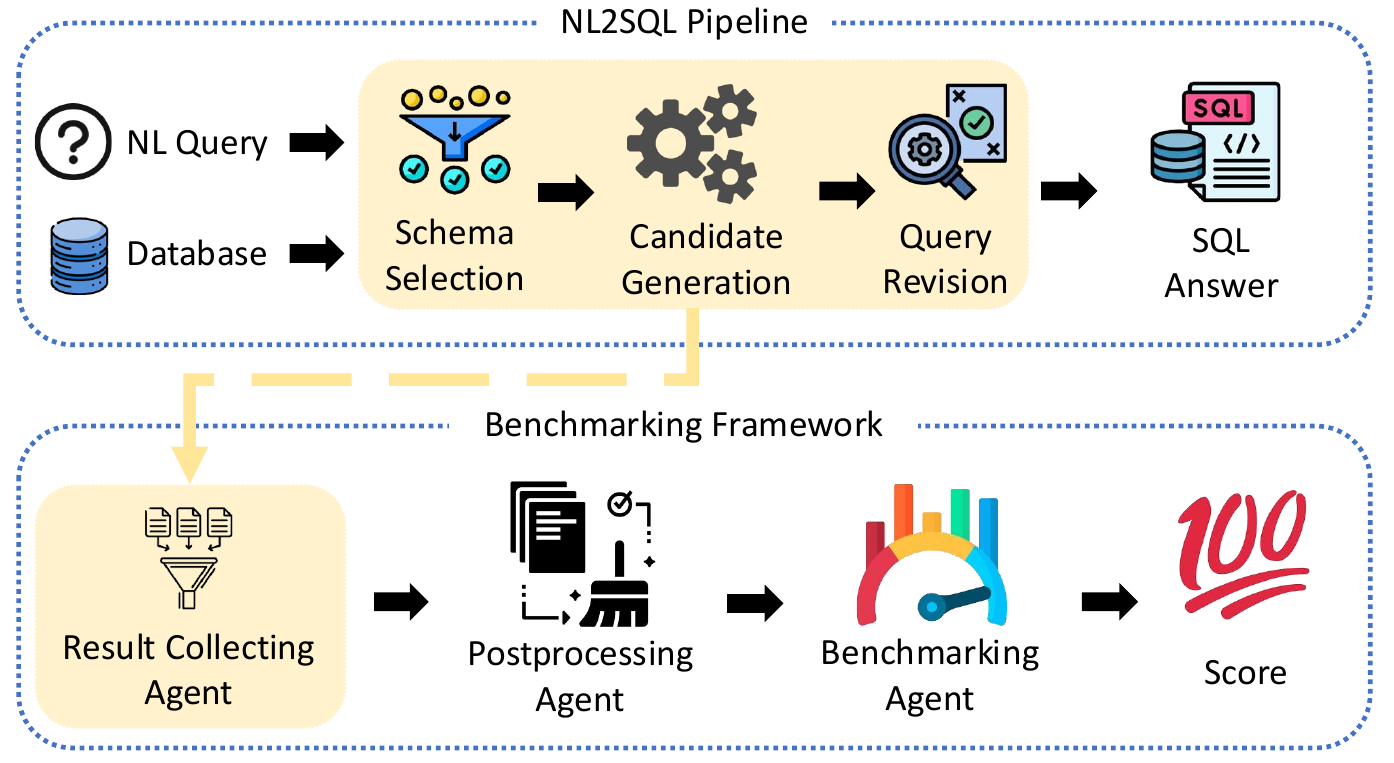}
  \caption{Overview of LLM-enabled NL2SQL approaches and our benchmarking framework.}
  \label{fig:benchmarking_framework}
\end{figure}

As articulated by~\citet{DBLP:journals/tkde/LiuSLMJZFLTL25}, these systems typically consist of three core modules: (1) \emph{Schema Selection}: select the most relevant tables and columns from the databases; (2) \emph{Candidate Generation}: generate candidate SQL queries based on the NL queries; (3) \emph{Query Revision}: refine the candidate SQL queries to generate the final SQL queries. These modules form the fundamental structure of most modern NL2SQL systems, as depicted in the top half of Figure \ref{fig:benchmarking_framework}.

The primary evaluation metrics currently employed in existing NL2SQL benchmarks~\cite{dataset-spider,dataset-bird,lei2025spider2}, typically employ three metrics, namely \emph{Exact Match (EM)}, \emph{Execution Accuracy (EX)}, and \emph{Valid Efficiency Score (VES)} that only concern the overall correctness and execution efficiency of the finally generated SQL. However, such metrics do not assess the effectiveness and efficiency of each core module individually, hindering deeper insights into where improvements are needed most urgently.

Furthermore, although an increasing number of solutions continue to shine on NL2SQL leaderboards~\cite{dataset-spider,dataset-bird,lei2025spider2}, the computational costs and latency incurred by LLMs remain overlooked. Additionally, variability in evaluation settings across studies complicates fair comparisons between different strategies for each module. There is no unified framework to perform a holistic comparison of the individual modules under the same experimental setting.

Therefore, the rapid development of LLM-based NL2SQL methods has outpaced systematic evaluation, leaving a critical gap in understanding their effectiveness and efficiency. These limitations highlight two urgent and interconnected research questions:
(1) \emph{How can we systematically measure the effectiveness and efficiency of individual NL2SQL modules?}, and (2) \emph{How can we establish a consistent, reproducible framework to evaluate these modules across different methods fairly?}
Answering these questions is crucial for understanding the performance of techniques used in each module, which ultimately guides the development of more accurate and efficient NL2SQL algorithms.

\textbf{Our work.}
To tackle these challenges, we present \texttt{\toolname}, the first modularized benchmarking framework specifically designed for systematic evaluation of the core modules \emph{Schema Selection}, \emph{Candidate Generation}, and \emph{Query Revision} of LLM-enabled NL2SQL solutions. We first review comprehensively different techniques implemented in each module by existing NL2SQL systems. To facilitate the modularized evaluation, we propose new evaluation metrics for each module, enabling a holistic assessment of the effectiveness and efficiency of these techniques. Then, we develop a flexible, multi-agent benchmarking framework consisting of three agents: \emph{Result Collecting}, \emph{Result Postprocessing}, and \emph{Benchmarking} agent. These agents are designed to be modular and configurable, allowing compatibility with diverse NL2SQL approaches.

We apply our framework to evaluate ten representative and open-sourced LLM-based NL2SQL approaches from the BIRD leaderboard\footnote{https://bird-bench.github.io/}~\cite{dataset-bird}, using two datasets, the BIRD \cite{dataset-bird} and the ScienceBenchmark \cite{dataset-sciencebenchmark} datasets, and two LLMs, DeepSeek-V3 \cite{deepseekv3r0324} and GPT-4o mini \cite{gpt4omini}. Then, we conduct a comprehensive analysis of their performance. Our comprehensive benchmarking analysis reveals that apart from the large room for improving accuracy at the modular level, computational efficiency is another obstacle to industrial adoption. Furthermore, our findings highlight previously underappreciated challenges regarding benchmark dataset quality and evaluation rules, areas that warrant immediate attention. Beyond just identifying these challenges, we further synthesize our insights into a series of practical guides and usability analyses (e.g., Sec. \ref{sec:schema_selection_practical_guide}, \ref{usability}), offering actionable guidance for developers to navigate the accuracy-efficiency trade-offs in real-world system development.

\textbf{Contributions.} We summarize our contributions as follows.

\begin{itemize}[nosep,noitemsep,left=1em]
  \item We comprehensively review the three core modules of LLM-enabled NL2SQL solutions, namely \emph{Schema Selection}, \emph{Candidate Generation}, and \emph{Query Revision}, on their details, representative implementations.

  \item We propose a set of new, fine-grained evaluation metrics
   specifically tailored for systematically assessing the effectiveness and efficiency of each individual module.
  Unlike existing overall metrics, our proposed metrics enable precise, detailed analyses of performance across multiple dimensions.


  \item
  We develop \texttt{\toolname}, the first modular, multi-agent benchmarking framework for evaluating core modules. Our framework comprises three seamlessly integrated agents—\emph{Result Collecting}, \emph{Result Postprocessing}, \emph{Benchmarking}—each designed with flexibility in mind, ensuring broad applicability across diverse NL2SQL methods.

  \item Leveraging \texttt{\toolname}, we comprehensively benchmark ten representative open-source NL2SQL approaches from the BIRD leaderboard. Our evaluation reveals substantial room for improvement in both accuracy and computational efficiency, warranting further study and providing valuable insights for researchers aiming to advance NL2SQL methodologies.

  \item Our in-depth analysis uncovers previously unnoticed yet critical limitations in the quality of existing datasets and evaluation rules. Our findings strongly advocate for the creation of more robust, reliable, and flexible benchmark datasets and evaluation methods in future research.


\item 
We distill the benchmarking results into practical guides and usability analyses, providing concrete recommendations and a systematic workflow to help practitioners effectively diagnose bottlenecks, navigate design trade-offs, and iteratively improve NL2SQL systems.
\end{itemize}

{\toolname} has been developed as a part of NeurDB\footnote{https://neurdb.com/}~\cite{NeurDBcidr, NeurDBSC} for supporting a friendly and yet effective query interface to the AI-powered autonomous data system.


%% file: sections/background.tex
\section{Comprehensive Review of LLM-enabled NL2SQL Solutions}
\label{sec:SurveyNL2SQL}

\subsection{NL2SQL Task Formulation}
\label{sec:nl2sql_formulation}

An NL2SQL task involves translating a natural language query into a corresponding SQL statement that can be executed on a relational database.
Specifically, given a natural language query \(Q\) on a database \(D\), the task is aimed at producing a SQL statement \(S\) based on \(Q\) and \(D\). Formally, the task can be defined as:
\(S = f\bigl(Q, D\bigr)\).

Most LLM-enabled NL2SQL approaches follow a three-stage architecture comprising three core modules: \emph{Schema Selection}, \emph{Candidate Generation}, and \emph{Query Revision}, which correspond to the three phrases \emph{pre-processing}, {query generation}, and {post-processing} \cite{DBLP:journals/tkde/LiuSLMJZFLTL25, shi2024surveyemployinglarge}, as depicted in Figure \ref{fig:benchmarking_framework}. Since these three modules all leverage LLMs to perform their functionalities, they need to consider various aspects such as how to design proper prompts, select few-shot examples, and decide the context length to feed into the LLMs. In the following, we comprehensively review how these modules are implemented in existing systems.

\subsection{Schema Selection Module}

Leveraging LLMs to generate SQL queries from an NL question requires constructing a prompt that incorporates the question, database schema, and task-specific instructions. However, the limited context window of LLMs poses challenges in accommodating large schemas~\cite{FloratouPZDHTCA24, lian2024chatbi}. Furthermore, lengthy prompts increase cost and degrade performance~\cite{liu-etal-2024-lost}.
Hence, the \emph{Schema Selection} module, which is also known as schema linking~\cite{DBLP:journals/tkde/LiuSLMJZFLTL25, lei-etal-2020-examining,wang-etal-2020-rat,li2023resdsql,cao2024rsl,pourreza2024din,caferouglu2024esql, guo-etal-2019-towards}, is implemented by many solutions to extract the most relevant tables and columns. We describe representative strategies employed in existing works below and summarize them in Table \ref{tab:schema_selection_classifiaction}.

\begin{table}[h]
  \centering
  \caption{Classification of Schema Selection strategies.}
  \renewcommand{\arraystretch}{0.9}
  \label{tab:schema_selection_classifiaction}
  \resizebox{0.8\columnwidth}{!}{%
  \setlength\tabcolsep{3pt}
  \begin{tabular}{lccc}
      \toprule
      Approach &
      \makecell{Few-shot\\CoT} &
      \makecell{Multi-Stage\\Pruning} &
      \makecell{Preliminary\\SQL} \\
      \midrule

      C3-SQL~\cite{dong2023c3} &  \No & \Yes & \No \\

      DIN-SQL~\cite{pourreza2024din} & \Yes & \No & \No \\

      MAC-SQL~\cite{macsql2025} & \Yes & \No & \No \\

      CHESS\textsubscript{(IR,SS,CG)}~\cite{talaei2024chess} & \Yes & \Yes & \No \\

      TA-SQL~\cite{qu2024before} & \No & \No & \Yes \\

      OpenSearch-SQL~\cite{xie2025opensearchsql} & \Yes & \Yes & \No \\

      RSL-SQL~\cite{cao2024rsl} & \Yes & \No & \Yes \\

      PET-SQL~\cite{li2024pet} & \Yes & \No & \Yes \\

      GSR~\cite{gsr} & \No & \No & \Yes \\
      \bottomrule
  \end{tabular}%
  }
\end{table}

\textbf{\emph{Few-Shot Chain-of-Thought (CoT).}}
This strategy utilizes the few-shot in-context learning~\cite{survey-in-context-learning, li-2023-practical} capabilities of LLMs supplemented by CoT prompting~\cite{wei2022chain, chu-etal-2024-navigate} to identify relevant tables and columns for NL queries, enabling LLMs to generalize without extensive task-specific training.
For instance, MAC-SQL~\cite{macsql2025} and DIN-SQL~\cite{pourreza2024din} select demonstrations that map NL queries to target schemas.
By guiding the LLM through a step-by-step process, the CoT method helps the model better interpret the NL query and effectively select relevant tables and columns.


\textbf{\emph{Multi-Stage Schema Pruning.}}
This strategy progressively narrows the database schema in multiple steps to ensure that only the necessary schemas are considered. For instance, CHESS\textsubscript{(IR,SS,CG)}~\cite{talaei2024chess} leverages locality-sensitive hashing and vector retrieval to efficiently identify relevant database values and catalogs, followed by a three-stage pruning process: broadly columns filtering, followed by selecting the most relevant table, and finally picking necessary columns. OpenSearch-SQL~\cite{xie2025opensearchsql} employs a similar way: information retrieval and column filtering, but chooses to recall relevant schema by leveraging both LLMs and vector retrieval.


\textbf{\emph{Preliminary-SQL Enhanced.}}
Instead of explicitly instructing LLMs on schema selection, this method begins by prompting the LLMs to generate a preliminary SQL query based on the natural language query. Once such a query is generated, relevant schema elements, such as tables, columns, and relationships, are extracted from it.
TA-SQL ~\cite{qu2024before} is a notable example of employing this strategy. It first generates a dummy SQL query and then extracts schema entities. PET-SQL~\cite{li2024pet} and GSR~\cite{gsr} also adopt similar methods.


\textbf{\emph{Hybrid Approaches.}}
Multiple strategies can be combined in a hybrid approach to improve their effectiveness. For example, RSL-SQL~\cite{cao2024rsl} combines zero-shot schema filtering and a preliminary SQL-based method, thus achieving a bidirectional schema linking design. CHESS\textsubscript{(IR,SS,CG)}~\cite{talaei2024chess} also employs a hybrid approach that combines few-shot CoT and schema pruning.

\textbf{Horizontal Comparison.} Few-shot CoT has low overhead and adapts quickly, but is sensitive to exemplar coverage and phrasing. Multi-stage pruning performs well and is robust under ambiguity, but it is very token-intensive and increases latency. Preliminary-SQL is token-efficient and precise when the draft is reliable, but degrades with draft errors or drift.

\subsection{Candidate Generation Module}
The Candidate Generation module is tasked with converting NL queries into candidate SQL statements.
This module directly interprets user intent and produces the corresponding SQL queries that align with the structure and constraints of the database schema.
We describe the representative strategies for this module as follows and present the classification of existing works in Table \ref{tab:candidate_generation}.

\textbf{\emph{Few-Shot CoT.}}
By explicitly incorporating intermediate reasoning steps in few-shot NL2SQL generation examples, few-shot CoT methods help LLMs process and solve complex logical processes in NL2SQL.
For example, DIN-SQL~\cite{pourreza2024din} employs human-designed CoT frameworks for different types of NL questions.

\textbf{\emph{Query Classification.}}
This strategy classifies NL queries into different classes according to their complexity and uses different prompts for each class.
DIN-SQL~\cite{pourreza2024din} is a representative example that employs a classification strategy to generate SQL candidates.

\textbf{\emph{Query Decomposition.}}
The decomposition-based strategy extends the CoT approach by decomposing a problem into smaller, manageable components~\cite{xie2024decomposition}.
The decomposer agent introduced by MAC-SQL~\cite{macsql2025} breaks the query into a series of intermediate steps, such as sub-questions, and generates corresponding sub-queries for each step before generating the final SQL.

\textbf{\emph{Intermediate Representation.}}
To bridge the gap between NL and SQL queries, intermediate representations (IRs)~\cite{guo-etal-2019-towards, NatSQL,li2023resdsql}, such as Pandas-like or SQL-like codes, have been introduced to facilitate the generation of SQL queries.
TA-SQL~\cite{qu2024before} employs Pandas-like code as an IR, DIN-SQL adopts the IR from NatSQL~\cite{NatSQL}, while OpenSearch-SQL invents an SQL-like language to encourage LLMs to focus more on logic before generating final SQL queries.

\textbf{\emph{Multiple Candidate Generation.}}
Some recent works choose to increase LLM calls or sampling numbers and generate multiple candidates before selecting the final answers among them. This strategy has been implemented in studies such as C3-SQL~\cite{dong2023c3}, CHESS~\cite{talaei2024chess}, MCS-SQL~\cite{lee2024mcs}, and OpenSearch-SQL~\cite{xie2025opensearchsql}.

\textbf{\emph{Hybrid Approaches.}}
Many works mentioned above also adopt multiple strategies to enhance performance.

\textbf{Horizontal Comparison.} Few-shot prompting is efficient but brittle to template and phrasing shift. Query classification improves alignment at modest cost but risks misrouting. Decomposition increases controllability and traceability but adds steps and propagates errors. Intermediate representations stabilize structure and reduce surface variance but may under-express rare constructs and add translation burden. Multi-candidate generation broadens coverage and hedges uncertainty but raises computation and depends on reliable selection.

\begin{table}[h]
\centering
\caption{Classification of Candidate Generation strategies.}
\renewcommand{\arraystretch}{1.0}
\setlength\tabcolsep{1pt}
\label{tab:candidate_generation}
\resizebox{0.45\textwidth}{!}{%
    \begin{tabular}{lccccc}
    \toprule
    Approach &
    \makecell{Few-shot\\CoT} &
    \makecell{Classifi\\cation} &
    \makecell{Decom\\position} &
    \makecell{Inter\\Represent} &
    \makecell{Multi\\Candidate} \\

    \midrule

    C3-SQL~\cite{dong2023c3} & \No & \No & \No & \No & \Yes \\

    DIN-SQL~\cite{pourreza2024din} & \Yes & \Yes & \No & NatSQL & \No \\

    MAC-SQL~\cite{macsql2025} & \Yes & \No & \Yes & \No & \No \\

    CHESS\textsubscript{(IR,SS,CG)}~\cite{talaei2024chess} & Zero-Shot & \No & \No & \No & \No \\

    CHESS\textsubscript{(IR,CG,UT)}~\cite{talaei2024chess} & Zero-Shot & \No & \No & \No & \Yes \\

    TA-SQL~\cite{qu2024before} & \No & \No & \No & Pandas-Like & \No \\

    OpenSearch-SQL~\cite{xie2025opensearchsql} & \Yes & \No & \No & SQL-Like & \Yes \\

    RSL-SQL~\cite{cao2024rsl} & Zero-Shot & \No & \No & \No & \Yes \\

    E-SQL~\cite{caferouglu2024esql} & \Yes & \No & \No & \No & \No \\

    PET-SQL~\cite{li2024pet} & \Yes & \No & \No & \No & \Yes \\
    \bottomrule
    \end{tabular}%
}
\end{table}

\subsection{Query Revision Module}
\label{sec:query_revision}

This module refines candidate SQL queries to produce the final query. Its tasks involve correcting SQL syntax errors, refining the query based on execution results, and selecting the best query from the multiple candidates generated. We survey the representative strategies as follows and summarize the classification of existing works based on these strategies in Table \ref{tab:revision_table}.

\textbf{\emph{LLM-Based.}}
Similar to Self-Refine~\cite{madaan2023selfrefine}, generally, the process begins by presenting the candidate SQL query to LLMs alongside the original natural language query. In a zero-shot setting, the model is explicitly instructed to review the SQL query for potential issues, such as syntax errors, semantic misalignments, or missing components, and produce a corrected SQL query.
DIN-SQL~\cite{pourreza2024din} employs this strategy effectively by integrating a self-correction module within its pipeline.

\textbf{\emph{Execution-Guided.}}
This strategy leverages the execution results of candidate SQL queries as a critical feedback mechanism to improve the accuracy. This process can be iterative, allowing multiple execution rounds, feedback analysis, and refinement until predefined termination conditions.
MAC-SQL~\cite{macsql2025} employs a dedicated Refiner Agent that automates the error detection and correction process. Similarly, CHESS ~\cite{talaei2024chess} starts with an initial draft query and executes it to gather feedback. The execution results, including error messages or outputs, are provided to the LLM, which adjusts and refines the SQL query accordingly.

\textbf{\emph{Consistency-Based.}}
Following the principle of self-consistency \cite{wang2022self}, this strategy generates multiple SQL queries through diverse reasoning paths, evaluates their execution results, and selects the most consistent query as the final output.
C3-SQL ~\cite{dong2023c3} incorporates a Consistency Output module, where multiple SQL queries are executed and filtered, and a voting mechanism is applied to the execution results to identify the most consistent query. CHESS ~\cite{talaei2024chess} selects the most consistent SQL query from three samples.

\textbf{\emph{Unit-Test-Based.}}
This strategy first generates multiple unit tests to highlight the differences between the candidate queries. Then, it selects the best query based on the evaluation results of the unit tests.
To the best of our knowledge, CHESS\textsubscript{(IR,CG,UT)} is the first approach that introduces this novel strategy.

\textbf{\emph{Question Rewriting.}} 
This strategy aims to better guide LLMs in revising and constructing SQL queries by enriching or rewriting the natural language questions. For example, E-SQL~\cite{caferouglu2024esql} enriches the questions by incorporating relevant database items, candidate predicates, and SQL generation steps. DART-SQL~\cite{dart-sql} utilizes database content to clarify and disambiguate questions through rewriting.

\textbf{\emph{Ranking and Selection.}}
Given multiple candidate queries, this strategy ranks the candidate queries based on some criteria and then selects the top-1 query from the query pool. A representative approach employing this method is CHASE-SQL~\cite{pourreza2025chasesql}, which fine-tunes a model specialized for ranking and selecting queries.

\textbf{\emph{Hybrid Approaches.}}
Similar to the Candidate Generation module, many works combine multiple of the above strategies to achieve better results. We present the detailed classification in Table \ref{tab:revision_table}.

\textbf{Horizontal Comparison.} Self-refine is lightweight but prone to superficial edits and regressions. Execution-guided repair addresses syntax and missing objects but adds iterations and cannot resolve silent semantic mismatches. Consistency-based selection reduces randomness and outliers but consumes sampling budget and may amplify shared biases. Unit-test-based checks deliver strong disambiguation but require specification effort and are sensitive to test quality. Question rewriting clarifies constraints and schema cues but can introduce bias or leakage. Ranking and selection consolidate decisions but may misrank under shift and add scoring overhead.

\begin{table}[h]
\centering
\caption{Classification of Query Revision strategies.}
\label{tab:revision_table}
\setlength\tabcolsep{1pt}
\renewcommand{\arraystretch}{1.0}
\resizebox{0.46\textwidth}{!}{%
\begin{tabular}{lcccccc}
    \toprule
    Approach &
    \makecell{LLM\\Based} &
    \makecell{Execution\\Guided} &
    \makecell{Consistency\\Based} &
    \makecell{Unit\\Test} &
    \makecell{Rank\\Select} &
    \makecell{Question\\Rewrite} \\
    \midrule

    C3-SQL~\cite{dong2023c3} & \No & Single-Turn & \Yes & \No & \No & \No \\

    DIN-SQL~\cite{pourreza2024din} & \Yes & \No & \No & \No & \No & \No \\

    MAC-SQL~\cite{macsql2025} & \No & Up to 3 & \No & \No & \No & \No \\

    CHESS\textsubscript{(IR,SS,CG)}~\cite{talaei2024chess} & \No & Single-Turn & \Yes & \No & \No & \No \\

    CHESS\textsubscript{(IR,CG,UT)}~\cite{talaei2024chess} & \No & Single-Turn & \No & \Yes & \No & \No \\


    OpenSearch-SQL~\cite{xie2025opensearchsql} & \No & Single-Turn & \Yes & \No & \No & \No \\

    RSL-SQL~\cite{cao2024rsl} & \No & Up to 5 & \No & \No & \Yes & \No \\

    E-SQL~\cite{caferouglu2024esql} & \No & \Yes & \No & \No  & \No & \Yes \\
    DART-SQL~\cite{dart-sql} & \No & \Yes & \No & \No  & \No & \Yes \\

    GSR~\cite{gsr} & \No & \Yes & \No & \No & \No & \No \\

    PET-SQL~\cite{li2024pet} & \No & Single-Turn & \Yes & \No & \No & \No \\

    CHASE-SQL~\cite{pourreza2025chasesql} & \No & \Yes & \No & \No & \Yes & \No \\
    \bottomrule
\end{tabular}%
}
\end{table}


%% file: sections/methodology.tex
\section{{\toolname}: Benchmarking and Evaluation Framework}
\label{sec:Methodology}

We now introduce {\toolname}, our modular benchmarking and evaluation framework for LLM-enabled NL2SQL approaches. We first propose fine-grained evaluation metrics tailored to each module, then present our multi-agent benchmarking framework for NL2SQL approaches, thus facilitating a comprehensive experimental analysis and providing deeper insights into their effectiveness, efficiency, and potential areas for improvement.

\subsection{Evaluation Metrics}
\label{sec:EvaluationMetrics}
We propose a set of metrics for each module, enabling a modularized and fine-grained evaluation of NL2SQL systems. By isolating the performance of individual modules, we facilitate a more nuanced understanding of their effectiveness and efficiency.

\subsubsection{Metrics for Schema Selection}
To evaluate the effectiveness of the Schema Selection module, we propose three metrics: \emph{Precision}, \emph{Recall}, and \emph{F1-score}, for both table and column retrieval. These metrics provide a standardized way of assessing the module's ability to select relevant schema elements while minimizing irrelevant selections.  Let \( R \) represent the set of relevant schema elements, and \( S \) represent the set of elements selected by the system. We define:
\begin{align}
\textrm{Precision}: \textrm{P} &= \lvert R \cap S \rvert \,/\, \lvert S \rvert \\
\textrm{Recall}: \textrm{R} &= \lvert R \cap S \rvert \,/\, \lvert R \rvert \\
\textrm{F1{-}score}: \textrm{F}_1 &= 2 \cdot P \times R \,/\, (P + R)
\end{align}


Precision measures the proportion of relevant elements retrieved, while Recall assesses the system’s ability to identify all relevant schema elements. A high F1-score reflects a balance between both. These metrics are calculated separately at the table and column levels to evaluate schema selection performance.

\subsubsection{Metrics for Candidate Generation}
Evaluating the effectiveness of this module is critical, as its output directly influences downstream processes such as query revision.
Existing benchmarking, such as Spider ~\cite{dataset-spider} and BIRD ~\cite{dataset-bird}, proposed the \emph{Execution Accuracy} (EX) metric as the proportion of examples for which the executed results of both the predicted and ground-truth SQLs are identical, relative to the overall number of test cases.

To provide a comprehensive evaluation, we propose more fine-grained metrics: \emph{Correct Rate} (CR)\footnote{The Correct Rate is equivalent to the Execution Accuracy used by Spider and BIRD.}, \emph{Incorrect Rate} (IR), and \emph{Error Rate} (ER), to evaluate a generated SQL query by assessing whether it produces (1) a Correct query (a correct result), (2) an Incorrect query (an incorrect result without runtime execution errors), or (3) an Error query (a runtime execution error), respectively.

Let $C$ be the set of correct SQL queries, $I$ be the set of Incorrect queries that execute without runtime errors, $E$ be the set of queries that produce execution errors, and $Q$ be the total number of queries generated. We define:
\begin{align}
\textrm{Correct Rate}: \textrm{CR} &= |C|\,/\,|Q| \\
\textrm{Incorrect Rate}: \textrm{IR} &= |I|\,/\,|Q| \\
\textrm{Error Rate}: \textrm{ER} &= |E|\,/\,|Q|
\end{align}

These metrics satisfy $C \cup I \cup E = Q$ and $C$, $I$, $E$ are pairwise disjoint, ensuring that each generated query falls into exactly one category and that $\textrm{CR} + \textrm{IR} + \textrm{ER} = 1$.

For SQL queries producing runtime execution errors, we classify them into the most common categories based on error messages and failure types such as \emph{no such table/column}, \emph{no such function}, \emph{syntax error}, \emph{timeout}, and \emph{other error types}.
This fine-grained categorization provides additional diagnostic insights into error analysis.

Furthermore, we adopt \texttt{Pass@k} for approaches producing multiple candidate queries, similar to CHESS~\cite{talaei2024chess}. Specifically, \texttt{Pass@k} measures the rate of producing at least one correct SQL query among \texttt{k} SQL queries produced per NL question. Formally,
\begin{equation}
\texttt{Pass@k} \;=\;
\frac{1}{N}\,
\sum_{i=1}^{N}
\mathbb{I}\Bigl[
\exists\, j\!\le\!k : \mathrm{Correct}\bigl(q_{i,j}\bigr)=1
\Bigr]
\label{eq:passatk}
\end{equation}

\noindent where the parameters are defined as follows.

\begin{itemize}[nosep,itemsep=1pt,left=1em]
  \item $N$: Total number of NL queries.
  \item $k$: Number of candidate SQL queries produced per NL query.
  \item $q_{i,j}$: The $j^{\text{th}}$ candidate query generated for the $i^{\text{th}}$ NL query.
  \item $\mathrm{Correct}(\cdot)$: Boolean function that returns $1$ if a candidate query yields the correct execution result, and $0$ otherwise.
  \item \(\mathbb{I}(\cdot)\): Indicator function: $1$ if the argument is true, $0$ otherwise.
\end{itemize}
This metric is mainly to measure the upper limit of the achievable accuracy when producing multiple candidates, thus providing important references for evaluating the Candidate Generation module and the Query Revision module.

\subsubsection{Metrics for Query Revision}
\label{sec:metric_query_revision}
To comprehensively evaluate the Query Revision module, we use the same primary metrics as for Candidate Generation: Correct Rate, Incorrect Rate, and Error Rate, with detailed error categorization. Additionally, we introduce new metrics: Correctness Improvement Rate (CI), Incorrect-to-Correct Rate (I2C), Error-to-Correct Rate (E2C), Correct-to-Incorrect Rate (C2I), and Correct-to-Error Rate (C2E), to specifically measure this module’s ability to enhance query quality and resolve errors without introducing regressions.

Let \( CR_{\text{pre}} \) and \( CR_{\text{post}} \) denote the Correct Rates before and after Query Revision. Similarly, let  \( C_{\text{pre}} \) and \( C_{\text{post}} \)  represent the Correct query sets, \( I_{\text{pre}} \) and \( I_{\text{post}} \) represent the Incorrect query sets, and \( E_{\text{pre}} \) and \( E_{\text{post}} \)  represent the Error query sets before and after Query Revision, respectively. Then, we define:
\begin{align}
\textrm{CI} &= (CR_{\text{post}}-CR_{\text{pre}})\,/\,CR_{\text{pre}} \\
\textrm{I2C} &= |I_{\text{pre}}\cap C_{\text{post}}|\,/\,|I_{\text{pre}}| \\
\textrm{E2C} &= |E_{\text{pre}}\cap C_{\text{post}}|\,/\,|E_{\text{pre}}| \\
\textrm{C2I} &= |C_{\text{pre}}\cap I_{\text{post}}|\,/\,|C_{\text{pre}}| \\
\textrm{C2E} &= |C_{\text{pre}}\cap E_{\text{post}}|\,/\,|C_{\text{pre}}|
\end{align}

Among the new metrics, CI, I2C, and E2C capture improvements in query quality, while C2I and C2E  identify potential regressions. Together with the primary metrics from the Candidate Generation module, these measures enable a comprehensive assessment of the module’s strengths and weaknesses.

\subsubsection{Efficiency Metrics}

Although an increasing number of novel solutions have achieved impressive performance on NL2SQL leaderboards, most overlook computational cost and latency, which are critical for real-world deployment~\cite{DBLP:journals/tkde/LiuSLMJZFLTL25}. This oversight significantly limits their deployment in real-world environments with strict constraints on resources, budgets, and real-time responsiveness.

To bridge this critical gap, we propose explicitly measuring the efficiency of individual NL2SQL modules using two metrics: \emph{\#Tokens} (the total number of tokens for prompts and completions) and \emph{\#LLM Calls} (the number of LLM invocations)\footnote{For the methods utilizing self-consistency on \(n\) outputs, we compute \emph{\#Tokens} and \emph{\#LLM Calls} by invoking the language model \(n\) times}. By systematically quantifying these metrics, our framework enables a detailed evaluation of computational overhead, 
fostering the development of NL2SQL solutions that are practically viable for widespread adoption.

\subsection{Multi-Agent Benchmarking Framework}
\label{sec:benchmarking_framework}

To facilitate the benchmarking, we develop a multi-agent benchmarking framework, as shown in the bottom half of Figure \ref{fig:benchmarking_framework}, comprising three specialized agents:

\emph{Result Collecting Agent.} This agent collects the results required for evaluation metrics from each module. It gathers selected schemas, candidate SQL queries, and refined SQL queries from the Schema Selection, Candidate Generation, and Query Revision modules, respectively. It also collects LLM usage statistics, such as token costs and the number of LLM calls. All the collected information is consolidated into structured JSON files. We provide an example of the collected JSON format in Appendix~\ref{appendix:json_structure_result_collection}.

\emph{Postprocessing Agent.}
This agent transforms raw outputs from upstream modules into a standardized format for benchmarking. Its main functions include extracting gold schema from gold SQL queries as a reference for schema selection accuracy, executing both candidate and refined SQL to compare results with gold SQL for correctness and error identification, and serializing evaluation outputs into structured JSON files for further analysis. Additionally, the agent compiles performance statistics, reporting total executions, success and failure counts, and per-question summaries.






\emph{Benchmarking Agent.}
Utilizing the processed datasets, this agent performs a comprehensive and modular evaluation of each module. Specifically, it leverages the previously defined evaluation metrics to compute quantitative performance indicators for each module. For each module, the agent aggregates relevant statistics and analyzes these metrics across varying difficulty levels of questions, enabling a granular assessment. The resulting evaluations are recorded in a structured format, providing detailed performance breakdowns that support comparative analysis across different strategies.

{\toolname} is designed for broad adaptability across NL2SQL systems, LLMs, and evaluation datasets. The \emph{Result Postprocessing} and \emph{Benchmarking} agents function identically for all NL2SQL methods, requiring no method-specific changes. Only the \emph{Result Collecting} agent adapts to each system, extracting and standardizing intermediate outputs with minimal wrapper code and configuration. {\toolname} allows users to switch LLMs or datasets via simple configuration without architectural modifications. Future updates will introduce an Orchestration agent to automate the evaluation workflow, enabling reproducible, end-to-end benchmarking of NL2SQL pipelines with minimal scripting effort.


%% file: sections/experiment.tex
\section{Experiments and Evaluation}
\label{sec:experiments}

We now present a comprehensive, modularized benchmarking of existing NL2SQL solutions, focusing on their three core modules, using our evaluation metrics and  {\toolname} framework to systematically and thoroughly analyze their effectiveness and efficiency.

\subsection{Experiment Settings}

\emph{Datasets.}
To ensure a rigorous and comprehensive evaluation of NL2SQL approaches, we utilize the BIRD dataset ~\cite{dataset-bird}, a large-scale and cross-domain dataset specifically designed to assess the capabilities of NL2SQL systems across diverse and complex scenarios, and the ScienceBenchmark dataset~\cite{dataset-sciencebenchmark}, which contains three real-world, highly domain-specific databases\footnote{The evaluation is conducted on the dev Set of BIRD and ScienceBenchmark, containing 1534 and 299 test cases, respectively.}.

\emph{Evaluated Approaches.}
We choose a list of representative and open-sourced approaches\footnote{Due to computational budget, we do not evaluate fine-tuned or RL-trained models.} from the BIRD leaderboard: C3-SQL~\cite{dong2023c3}, DIN-SQL~\cite{pourreza2024din}, MAC-SQL~\cite{macsql2025}, TA-SQL~\cite{qu2024before}, CHESS\textsubscript{(IR,SS,CG)}~\cite{talaei2024chess}, CHESS\textsubscript{(IR,CG,UT)}~\cite{talaei2024chess}, E-SQL~\cite{caferouglu2024esql}, RSL-SQL~\cite{cao2024rsl}, OpenSearch-SQL~\cite{xie2025opensearchsql} and GSR~\cite{gsr}.
These approaches cover diverse strategies for the three core modules mentioned previously.

\emph{Language Models.}
To ensure a fair and consistent comparison across all approaches while minimizing the experimental costs, we utilize DeepSeek-V3~\cite{deepseekv3r0324} and GPT-4o-mini~\cite{gpt4omini} as the unified LLMs for all evaluations\footnote{We use the DeepSeek-V3-0324 Release version.}.
For model-related hyperparameters such as temperature, maximum tokens, and prompts, we retain the original configurations used by each evaluated method to ensure objectivity in evaluation.
For the approaches that require models for embedding and retrieval, we choose the model \texttt{bge-large-en-v1.5}~\cite{bge_embedding}.

\subsection{Results for Schema Selection Module}

\begin{table}[h]
  \centering
  \caption{Analysis results of Schema Selection on BIRD using DeepSeek-V3. The best results are in \fst{bold and underlined} while the second-best results are \snd{underlined}.}
  \setlength\tabcolsep{1pt}
  \small
  \renewcommand{\arraystretch}{1.0}
  \label{tab:schema_selection_bird_deepseek}
  \resizebox{\columnwidth}{!}{%
  \begin{tabular}{l|c|c|c|c|c|c|c|c}
     \toprule
        \multirow{2}{*}{\makecell{\\Approach}}   & \multicolumn{3}{c|}{Table Selection}  & \multicolumn{3}{c|}{Column Selection}  & \multicolumn{2}{c}{Efficiency} \\ \cline{2-9}
        &  \makecell{Precis-\\ion (\%)}  &  \makecell{Recall\\(\%)} &  \makecell{F1-score\\(\%)} &  \makecell{Precis-\\ion (\%)} &  \makecell{Recall\\(\%)} &  \makecell{F1-score\\(\%)} &  \makecell{\#Tok-\\ens} &  \makecell{\#LLM\\calls}\\
        \midrule

        C3-SQL                & 50.03  & 98.80  & 64.63  & 27.88  & 95.03  & 41.76  & 15886  & 20   \\
        DIN-SQL               & 93.27  & 95.47  & 93.41  & \snd{89.10}  & 88.76  & \fst{88.01}  & 7360   & \fst{1}   \\
        MAC-SQL               & 32.74  & \fst{99.84}  & 46.70  & 15.84  & \fst{98.70}  & 26.12  & \fst{3179}   & \fst{1}   \\
        CHESS\textsubscript{(IR,SS,CG)} & \fst{94.50}  & 95.85  & \fst{94.35}  & \fst{89.78}  & 88.00  & \snd{87.66}  & 307894   & 78.56   \\
        TA-SQL                & \snd{93.66}  & 95.73  & \snd{93.77}  & 82.46  & 90.77  & 84.89  & \snd{4249}   & \fst{1}   \\
        RSL-SQL               & 88.53  & 97.09  & 91.21  & 55.03  & 93.01  & 63.37  & 5790   & \fst{1}  \\
        OpenSearch-SQL        & 32.80  & \snd{99.62}  & 46.72  & 16.97  & \snd{97.35}  & 27.49  & 6698   & \snd{3}   \\
     \bottomrule
  \end{tabular}%
  }
\end{table}

\subsubsection{Overall performance.}

Table~\ref{tab:schema_selection_bird_deepseek} reports the overall results of the schema selection module on the BIRD with DeepSeek-V3. The results of the remaining three settings are deferred to Appendix~\ref{appendix:schema_selection}. Across all four settings, we observe highly consistent rankings.

\textbf{Table selection.} As shown in Table~\ref{tab:schema_selection_bird_deepseek}, CHESS\textsubscript{(IR,SS,CG)} and TA-SQL achieved the top two F1 scores, followed closely by \textsc{DIN-SQL} and \textsc{RSL-SQL}. This pattern largely persists when swapping the LLM and switching to ScienceBenchmark, with minor shifts. These consistent outcomes show the effectiveness of their Multi-Stage Schema Pruning (as in CHESS\textsubscript{(IR,SS,CG)}) and Preliminary-SQL Enhanced strategies (as in \textsc{TA-SQL}). Few-Shot CoT (\textsc{DIN-SQL}) also delivers strong, often third-best F1, suggesting that lightweight reasoning with carefully curated exemplars can narrow the gap to heavier multi-stage pipelines.

\textbf{Column selection.} On BIRD, \textsc{DIN-SQL} leads the column-level F1, with \textsc{CHESS} and \textsc{TA-SQL} close behind; On ScienceBenchmark, leadership alternates among \textsc{CHESS} and \textsc{TA-SQL}, with \textsc{DIN-SQL} remaining competitive. Compared to BIRD, ScienceBenchmark tends to depress column-selection precision, leading to lower column-F1 across systems. We conjecture this stems from the fact that the schema complexity of ScienceBenchmark datasets is higher than that of BIRD.
Indeed, methods such as \textsc{MAC-SQL} and \textsc{OpenSearch-SQL} often attain very high recall at both table and column levels but suffer from low precision.

\textit{\textbf{Insight 1}: Multi-Stage Schema Pruning, Pre-SQL Enhanced strategies, and Few-Shot CoT reasoning are consistently effective for both table and column selection. However, their gains depend on careful implementation details. The predominant issue is over-selection, especially on complex schemas (ScienceBenchmark). This suggests adding precision-oriented controls to curtail long tails of spurious columns.}

\textit{\textbf{Insight 2:} Relative rankings are stable across LLMs and datasets, with only mild re-ordering at the top. This robustness indicates the observed advantages stem from algorithmic choices rather than opportunistic synergy with a particular LLM or dataset.}

\subsubsection{Cost efficiency and performance on large-scale databases.}
Across settings, we observe a cost-quality frontier. \textsc{CHESS}\textsubscript{(IR,SS,CG)} attains top table/column selection but incurs substantial token cost, and LLM calls pose significant barriers to deployment, where latency and cost matter. By contrast, \textsc{TA-SQL} and \textsc{DIN-SQL} deliver near-top F1 with a single call and low tokens, offering a favorable accuracy–efficiency trade-off.

These trends persist on ScienceBenchmark, which better approximates real-world schema density than academic suites such as Spider~\cite{dataset-spider} and BIRD~\cite{dataset-bird}. On ScienceBenchmark we consistently see lower column-selection precision and stronger over-selection pressure. From a scalability standpoint, schema selection cost often grows with the number of tables/columns included in prompts or retrieval stages. This underscores the need for future research into lightweight pruning strategies that maintain accuracy while minimizing computational overhead.

\textit{\textbf{Insight 3:}  Multi-stage pipelines (e.g., \textsc{CHESS}) yield the highest accuracy but at high token/call budgets; single-call pre-SQL/CoT approaches (e.g., \textsc{TA-SQL}, \textsc{DIN-SQL}) provide the strongest production-ready trade-off. ScienceBenchmark stresses these dynamics with denser schemas, revealing precision-cost trade-off as a crucial concern for maintaining accuracy under strict cost/latency constraints.}

\begin{figure}[htbp]
  \centering
  \includegraphics[width=0.8\linewidth]{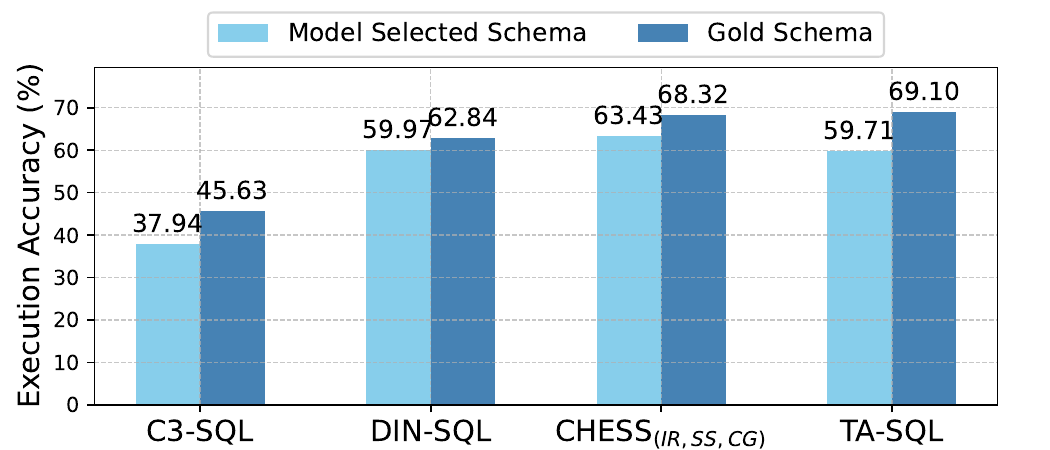}
  \caption{Execution accuracy using different schemas.}
  \label{fig:gold_schema}
\end{figure}

\subsubsection{Influence of schema selection on overall performance}
\label{sec:influence_schema_selection}
To assess the influence of accurate schema selection on the overall performance, we conducted experiments comparing execution accuracies using model-selected schemas versus gold schemas on some approaches\footnote{Due to budget limitations, we only chose C3-SQL, DIN-SQL, CHESS, and TA-SQL.} using the BIRD dev set and Deepseek-V3. Figure~\ref{fig:gold_schema} presents the execution accuracies for each approach under both conditions. The results indicate that the use of gold schemas significantly enhances execution accuracy across all evaluated approaches. Specifically, for C3-SQL, the execution accuracy increased from 37.94\% to 45.63\%, a significant improvement of 20.27\% relatively. TA-SQL's execution accuracy improved from 59.71\% to 69.10\%, the highest increase among all approaches, with a gain of 15.73\%.
This observation confirms the critical role of accurate schema selection -- the provision of the correct schemas to the subsequent modules eliminates errors arising from incorrect schema information, which is aligned with the observation of~\cite{macsql2025, FloratouPZDHTCA24}.
While some recent studies~\cite{maamari2024death,caferouglu2024esql} suggest that schema selection is unnecessary with today’s strongest LLMs, we contend it remains indispensable for robust NL2SQL systems. Not only are cutting-edge models often inaccessible under strict compute or budget constraints, but feeding entire enterprise schemas, potentially comprising thousands of tables and columns, can exceed model context windows, increase token costs, and ultimately degrade output quality\cite{liu-etal-2024-lost}.

\textit{\textbf{Insight 4:} Accurate schema selection is critical in improving execution accuracy and preventing error propagation, and remains essential for robust real-world deployment due to practical constraints.}

\subsubsection{Practical guide.}
\label{sec:schema_selection_practical_guide}

We recommend implementing a two-stage schema pruning pipeline that balances effectiveness and efficiency for schema selection. In the first stage, employ lightweight retrieval methods (e.g., embedding-based similarity search or keyword matching) to broadly filter irrelevant tables and columns, prioritizing high recall to minimize the risk of excluding relevant schema elements. In the second stage, leverage LLMs with carefully designed prompts to perform precision-oriented refinement on the reduced schema space, focusing on eliminating irrelevant elements that would otherwise increase token costs and degrade downstream performance.

Additionally, for production systems handling ultra-large databases, we recommend maintaining a schema cache indexed by query embeddings to avoid redundant schema selection for similar queries, and implementing adaptive schema expansion that dynamically adjusts the schema scope based on initial query execution feedback.

For resource-constrained environments, Preliminary-SQL Enhanced and Few-shot CoT strategies are favored, which leverage the inherent generation capabilities of LLMs without multiple calls, offering a practical balance between accuracy and efficiency.

\subsection{Results for Candidate Generation Module}
\label{sec:result_candidate_generation}

\begin{table}[t]
  \centering
  \caption{Results of Candidate Generation on BIRD dev with two LLMs. The best results inside each block are \fst{bold+underlined}, the second-best are \snd{underlined}. We use \texttt{Pass@1} for methods producing multiple candidates.}
  \scriptsize
  \renewcommand{\arraystretch}{1.0}
  \setlength\tabcolsep{1pt}
  \label{tab:candidate_generation_bird_both_llms}
  \resizebox{\columnwidth}{!}{%
  \begin{tabular}{l|ccc|cc|ccc|cc}
    \toprule
    \multirow{2}{*}{\makecell{\\Approach}}
    & \multicolumn{5}{c|}{\textbf{DeepSeek-V3}}
    & \multicolumn{5}{c}{\textbf{GPT-4o-mini}} \\
    \cline{2-11}
    & \makecell{CR\\(\%)} & \makecell{IR\\(\%)} & \makecell{ER\\(\%)} & \makecell{\#Tok.} & \makecell{\#Calls}
    & \makecell{CR\\(\%)} & \makecell{IR\\(\%)} & \makecell{ER\\(\%)} & \makecell{\#Tok.} & \makecell{\#Calls} \\
    \midrule
    C3-SQL & 38.14 & 50.07 & 11.80 & 11786   & 20
           & 35.66 & 51.83 & 12.52 & \snd{1353} & 20 \\
    DIN-SQL & 59.58 & 35.52 & 4.89  & 6680    & \snd{2}
            & \snd{56.65} & 34.49 & 8.87  & 6523 & \snd{2} \\
    MAC-SQL & 58.74 & 33.83 & 7.43  & 2829    & \fst{1}
            & 52.74 & 35.92 & 11.34 & 3031 & \fst{1} \\
    CHESS\textsubscript{(IR,SS,CG)}
            & 59.78 & \fst{33.12} & 7.11  & \fst{1017} & \fst{1}
            & 53.42 & 41.03 & \fst{5.54} & \fst{934} & \fst{1} \\
    CHESS\textsubscript{(IR,CG,UT)}
            & \snd{62.91} & \snd{33.12} & \snd{3.98} & 234394  & 20
            & 55.08 & \snd{32.01} & 12.91 & 228177 & 20 \\
    TA-SQL  & 59.71 & 34.42 & 5.87  & \snd{2592}  & \snd{2}
            & 53.06 & 33.57 & 13.36 & 2427 & \snd{2} \\
    GSR     & 57.37 & 37.09 & 5.48  & 3846   & \snd{2}
            & 52.67 & 40.29 & \snd{7.04} & 2006 & \snd{2} \\
    E-SQL   & 57.50 & 35.07 & 7.43  & 13118  & \fst{1}
            & \fst{58.02} & \snd{32.07} & 9.91  & 13031 & \fst{1} \\
    RSL-SQL & 59.00 & 33.38 & 7.62  & 11370  & 5
            & 55.54 & 35.01 & 9.45  & 2219 & 5 \\
    OpenSearch-SQL
            & \fst{66.10} & \fst{30.83} & \fst{3.06} & 100095 & 21
            & 56.52 & 35.33 & 8.15  & 8984 & 21 \\
    \bottomrule
  \end{tabular}%
  }
\end{table}

\begin{table}[t]
  \centering
  \caption{Results of Candidate Generation on ScienceBenchmark. The best results are \fst{bold+underlined} and the second-best are \snd{underlined}. We report \texttt{Pass@1} for methods producing multiple candidates.}
  \scriptsize
  \renewcommand{\arraystretch}{1.0}
  \setlength\tabcolsep{1.2pt}
  \label{tab:candidate_generation_scibench_both_llms}
  \resizebox{\columnwidth}{!}{%
  \begin{tabular}{l|ccc|cc|ccc|cc}
    \toprule
    \multirow{2}{*}{\makecell{\\Approach}}
      & \multicolumn{5}{c|}{\textbf{DeepSeek-V3}}
      & \multicolumn{5}{c}{\textbf{GPT-4o-mini}} \\
    \cline{2-11}
      & \makecell{CR\\(\%)} & \makecell{IR\\(\%)} & \makecell{ER\\(\%)} & \makecell{\#Tok.} & \makecell{\#Calls}
      & \makecell{CR\\(\%)} & \makecell{IR\\(\%)} & \makecell{ER\\(\%)} & \makecell{\#Tok.} & \makecell{\#Calls} \\
    \midrule
    C3-SQL & 54.18 & 42.81 & 3.01  & 12110 & 20
           & \snd{50.84} & 42.81 & 6.35  & 1350 & 20 \\
    DIN-SQL & 52.84 & 44.82 & \snd{2.34} & 6149  & \snd{2}
            & 42.81 & 48.49 & 8.70  & 5932 & \snd{2} \\
    MAC-SQL & 47.16 & 42.81 & 10.03 & 3514  & \fst{1}
            & 44.15 & \snd{41.81} & 14.05 & 3734 & \fst{1} \\
    CHESS\textsubscript{(IR,SS,CG)}
            & 37.92 & 55.37 & 6.71  & \snd{978} & \fst{1}
            & \fst{53.42} & \fst{41.03} & \fst{5.54} & \snd{934} & \fst{1} \\
    CHESS\textsubscript{(IR,CG,UT)}
            & \snd{54.85} & 39.46 & 5.69  & 259569 & 20
            & 40.80 & 46.82 & 12.37 & 247969 & 20 \\
    TA-SQL  & \fst{59.53} & \snd{38.46} & \fst{2.01} & 2712  & \snd{2}
            & 50.17 & 43.81 & \snd{6.02} & 2523 & \snd{2} \\
    GSR     & 40.47 & 47.49 & 12.04 & \fst{824} & \snd{2}
            & 40.47 & 52.17 & 7.36  & \fst{766} & \snd{2} \\
    E-SQL   & 54.52 & 40.47 & 5.02  & 6542  & \fst{1}
            & 38.80 & 45.82 & 15.38 & 12736 & \fst{1} \\
    RSL-SQL & 45.48 & 39.46 & 15.05 & 2877  &  5
            & 40.47 & 52.84 & 6.69  & 1532 &  5 \\
    OpenSearch-SQL
            & 38.46 & \fst{27.09} & 34.45 & 103392 & 21
            & 42.14 & 46.82 & 11.04 & 8355  & 21 \\
    \bottomrule
  \end{tabular}%
  }
\end{table}

\subsubsection{Overall performance.}
\label{sec:candidate_generation_overall_performance}
We evaluate \texttt{Pass@1} for comparability. Table~\ref{tab:candidate_generation_bird_both_llms} and Table~\ref{tab:candidate_generation_scibench_both_llms} show that, unlike schema selection, candidate generation is highly sensitive to both the dataset and the LLM—rankings shift, and there is no universal winner.

\textbf{BIRD}. With DeepSeek-V3, OpenSearch-SQL attains the best Correct rate (CR) with the lowest Incorrect Rate (IR), and \textsc{CHESS}\textsubscript{(IR,CG,UT)} is a close second. With GPT-4o-mini, the ordering changes: \textsc{E-SQL} achieves the highest CR, and simpler pipelines such as \textsc{DIN-SQL} and \textsc{TA-SQL} remain competitive at low token budgets. These shifts indicate that candidate generation quality depends strongly on the LLM; the same pipeline can rank differently when the LLM changes.

\textbf{ScienceBenchmark}. With DeepSeek-V3, \textsc{TA-SQL} yields the strongest CR at low Error Rate (ER), while breadth-oriented systems like OpenSearch-SQL see degraded CR due to increased fragility on complex schemas. With GPT-4o-mini, \textsc{CHESS}\textsubscript{(IR,SS,CG)} becomes the most effective at \texttt{Pass@1} with \emph{one} call and sub-1K tokens.

\textbf{Error profile}. Across all four blocks, the dominant failure mode is IR: syntactically valid but \emph{semantically misaligned} SQL substantially outnumbers execution errors. Typical IR patterns include (i) wrong join path or missing bridge table, (ii) predicate misalignment, (iii) aggregation/GROUP BY/HAVING mismatches, and (iv) projection-level omissions or extra columns. ER is non-negligible for some complex databases on ScienceBenchmark, but even there IR constitutes the majority of errors; improving \emph{semantic alignment} is therefore the primary lever for improving the accuracy.

\textbf{Cost–efficiency}. A clear cost–quality frontier emerges. Multi-candidate pipelines (e.g., OpenSearch-SQL, \textsc{CHESS}\textsubscript{(IR,CG,UT)}) can lead on BIRD but often require tens of calls and very high token budgets; on ScienceBenchmark, their advantage weakens. In contrast, token-efficient pipelines (e.g., \textsc{CHESS}\textsubscript{(IR,SS,CG)}, \textsc{TA-SQL}, sometimes \textsc{DIN-SQL}) achieve competitive CR at a fraction of the cost.

\textit{\textbf{Insight 5:} The dominant failure mode is semantic mismatch. Adding semantic checks that align the SQL with the question will be helpful. Results shift between LLMs and datasets, indicating that performing evaluation on targeting databases and backbone LLMs before adopting specific NL2SQL approaches is necessary.}

We also conducted a detailed error analysis and result analysis on different difficulty levels of questions. The full results are presented in Appendix~\ref{appendix:error_analysis_candidate_generation}.

\subsubsection{\texttt{Pass@k} evaluation for multiple candidates.}
\label{sec:pass_k_eval}

\begin{figure}[h]
  \centering
  \includegraphics[width=1.0\linewidth]{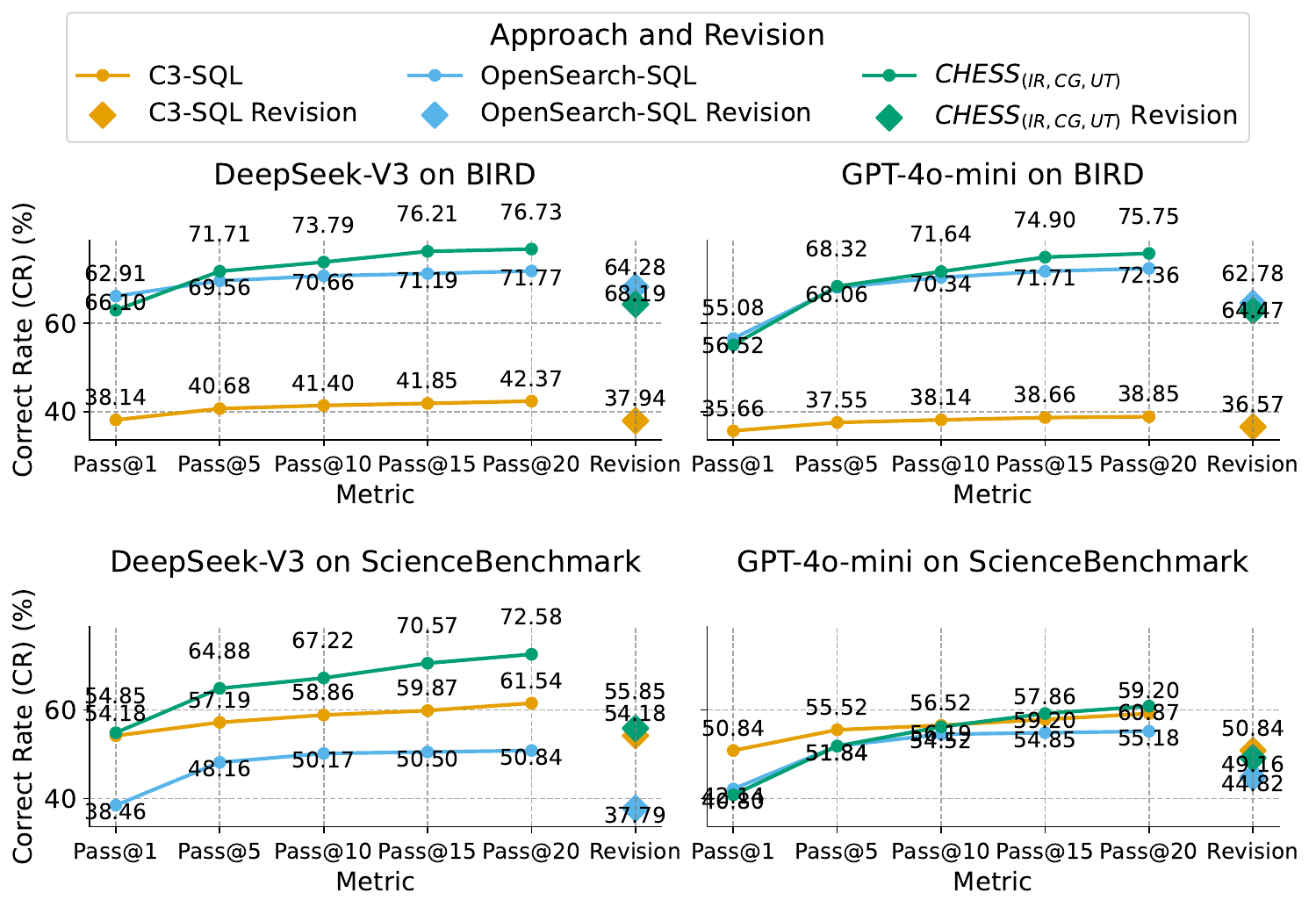}
  \caption{\texttt{Pass@k} results for multiple-candidate approaches.}
  \label{fig:pass_at_k_revision}
\end{figure}

To evaluate methods that produce multiple candidate queries, we report \texttt{Pass@k} for $k \in \{1, 5, 10, 15, 20\}$ across four settings: BIRD with DeepSeek-V3 and GPT-4o-mini, and ScienceBenchmark with DeepSeek-V3 and GPT-4o-mini. As illustrated in Figure \ref{fig:pass_at_k_revision}, across all subfigures, the curves show the same shape: execution accuracy rises with larger $k$, confirming the benefit of generating multiple candidates. The marginal gains, however, diminish as $k$ increases, indicating limited returns beyond a moderate number of candidates. The consistency among all subfigures indicates that our conclusion is robust to both the backbone LLM and the evaluation dataset. Since \texttt{Pass@k} represents an upper bound on achievable accuracy, these results suggest that a well-designed Query Revision module, capable of reliably selecting the optimal query, has the potential to elevate the accuracy close to this theoretical limit.

\textit{\textbf{Insight 6:} Increasing the number of candidates raises potential accuracy across all LLMs/datasets, but the gains diminish rapidly; if the Query Revision module can consistently select the best query from the candidates, overall accuracy can nearly reach the upper bound.}

\subsubsection{Practical guide.}
\label{sec:candidate_generation_practical_guide}
Our evaluation demonstrates that syntactically valid but semantically misaligned queries constitute the dominant failure case, far exceeding execution errors. We recommend implementing execution-result-based semantic validation during the generation phase rather than deferring all validation to the revision stage. Specifically, for each generated candidate query, execute it against the database and analyze the result schema (column names, data types) and sample result values to detect potential semantic mismatches. This early detection mechanism can prevent the propagation of semantic errors to downstream modules and reduce the burden on query revision.

For systems generating multiple candidates, our \texttt{Pass@k} analysis reveals that increasing diversity ($k$ > 10) yields diminishing returns; instead, practitioners should focus on improving the quality of top-ranked candidates through better prompt engineering and schema-aware generation strategies. Intermediate Representations (e.g., NatSQL, Pandas-like code) show promise in improving correctness, but introduce syntax translation overhead; practitioners should carefully weigh this trade-off based on their target SQL dialect complexity. In addition, using early-exit and expanding to larger $k$ only for high-uncertainty queries is a more cost-efficient approach.

For production deployment, we recommend implementing progressive generation that starts with simpler query structures and incrementally adds complexity only when necessary, as our difficulty-level analysis shows that accuracy degrades sharply for challenging queries, suggesting that avoiding unnecessary complexity can improve overall robustness.

\subsection{Results for Query Revision Module}
\label{sec:result_query_revision}

\begin{figure}[htbp]
  \centering
  \includegraphics[width=1.0\linewidth]{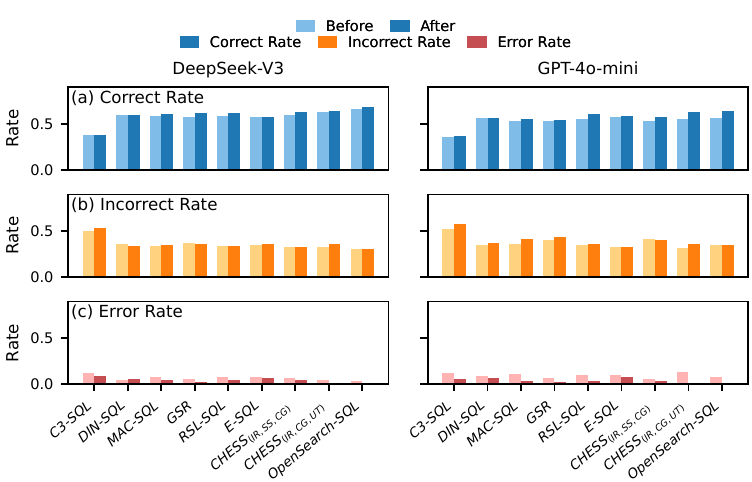}
  \caption{Analysis of the Query Revision module on BIRD.}
  \label{fig:revision_overall_bird}
\end{figure}

\begin{figure}[htbp]
  \centering
  \includegraphics[width=1.0\linewidth]{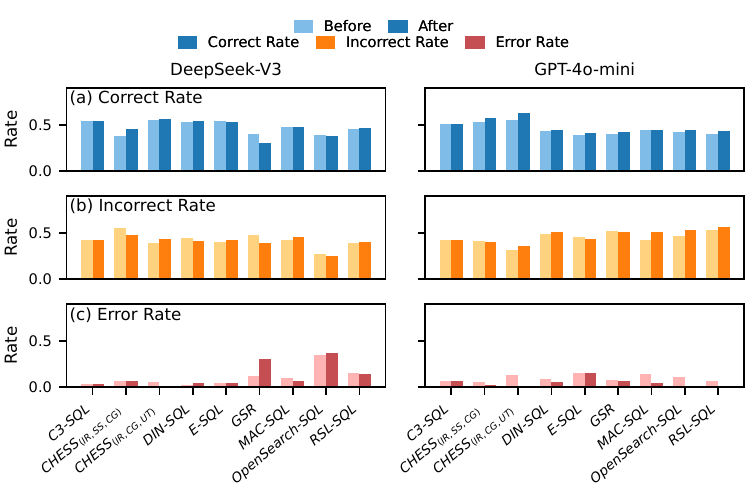}
  \caption{Analysis of the Query Revision module on ScienceBenchmark.}
  \label{fig:revision_overall_sciencebenchmark}
\end{figure}

\subsubsection{Overall performance.}
Figures \ref{fig:revision_overall_bird} and \ref{fig:revision_overall_sciencebenchmark} compare before vs. after the Revision stage on Correct Rate (CR), Incorrect Rate (IR), and Error Rate (ER) for all approaches.

\textbf{BIRD.} Across most systems, Revision raises CR while reducing IR and keeping ER low. The gains are largest for pipelines that already provide reasonably good candidates: Revision acts mainly as a selector/repair step that fixes small semantic gaps and filters poor candidates. Methods that rank/select the best SQL query from multiple SQL candidates also improve, but their CR uplift depends on the revision policy’s ability to reliably select the optimal query.

\textbf{ScienceBenchmark.} The improvements are more heterogeneous across systems. Some systems still obtain clear CR gains, but others see smaller gains or even a mild rise in IR. Furthermore, the extent of performance improvement varies depending on the backbone LLM employed, even when using the same NL2SQL approach. The divergence can be attributed to two compounding factors. First, the ScienceBenchmark dataset generally features more complex schemas and more challenging questions than the BIRD dataset, resulting in a more demanding evaluation of query revision strategies. Second, the query revision strategies of many NL2SQL approaches are, to some extent, specifically tailored to the BIRD dataset and therefore do not fully generalize to the ScienceBenchmark scenarios, struggling to address the unique challenges that the latter presents.

\textbf{Error profile.} In all settings, the predominant failure type after revision is still IR (syntactically valid yet semantically misaligned SQL), which greatly outnumbers ER.
Therefore, the upper bound for further improvements in CR primarily depends on mitigating IR, while reducing ER is still necessary.

\begin{figure}[htbp]
  \centering
    \begin{subfigure}[b]{0.7\linewidth}
    \centering
    \includegraphics[width=\linewidth]{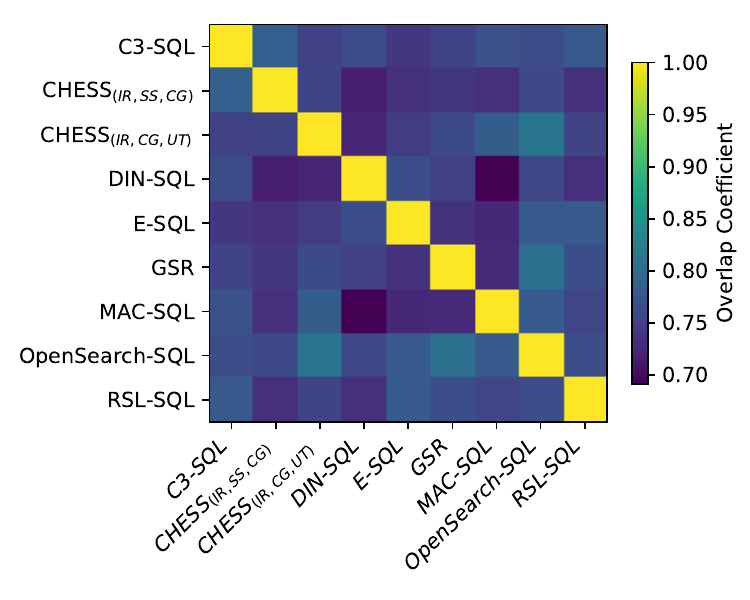}
    \caption{Overlap on the BIRD (using DeepSeek-V3)}
    \label{fig:overlap_a}
  \end{subfigure}
  \hfill
  \begin{subfigure}[b]{0.7\linewidth}
    \centering
    \includegraphics[width=\linewidth]{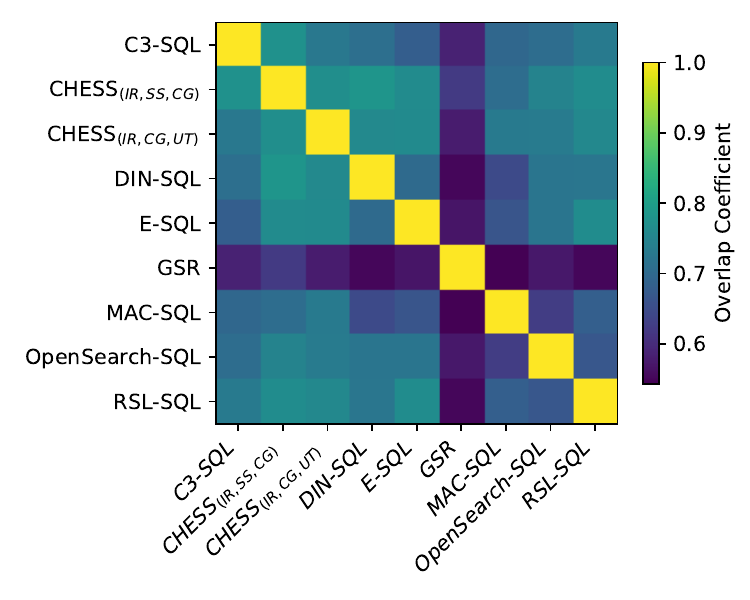}
    \caption{Overlap on the ScienceBenchmark (using DeepSeek-V3)}
    \label{fig:overlap_b}
  \end{subfigure}
  \caption{The coefficient heatmap of different solutions on Incorrect cases.}
  \label{fig:incorrect_overlap}
\end{figure}

\begin{figure}[htbp]
  \centering
  \includegraphics[width=0.8\linewidth]{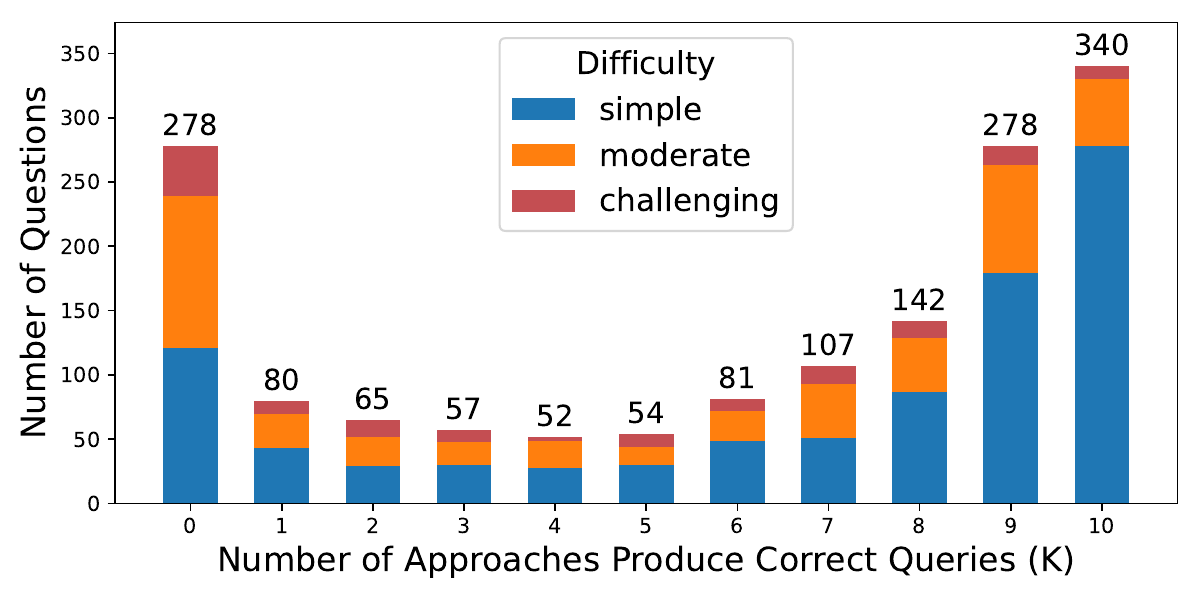}
  \caption{Breakdown of questions by number of approaches producing Correct queries (for different difficulty levels).}
  \label{fig:revision_correct_distribution}
\end{figure}

\subsubsection{An in-depth analysis of incorrect cases.}
\label{sec:incorrect_overlap}
Through an in-depth analysis of incorrect cases, we observed an unexpectedly high degree of overlap in incorrect queries generated by different methods using DeepSeek-V3, as illustrated in Figure \ref{fig:incorrect_overlap}. Similar results using GPT-4o-mini are provided in Appendix~\ref{appendix:revision_in_depth_analysis}. Furthermore, as shown in Figure \ref{fig:revision_correct_distribution}, a significant portion -- 278 out of the total 1,534 questions from the BIRD Dev Set -- could not be correctly solved by any of the evaluated approaches (the first vertical bar, which indicates no approach can produce correct SQL queries). This substantial overlap suggests a shared set of underlying limitations across existing NL2SQL methods, potentially involving insufficient semantic understanding, inadequate query generation strategies, or ineffective query revision mechanisms.

To uncover the root causes behind this, we conducted an additional human evaluation of these 278 problematic questions. Remarkably, the results revealed that a considerable fraction of these questions were paired with inaccurate gold SQL queries, which is aligned with the observation of~\cite{liu2025nl2sql-bugs}. This discovery highlights the often overlooked but critical issue of benchmark dataset quality and raises concerns that current assessments of NL2SQL methods on the BIRD dataset may have been compromised or misled by inaccuracies inherent in the benchmark. For a more detailed analysis of the BIRD dataset, please see Section \ref{sec:datasets_analysis}.

\textit{\textbf{Insight 7:} Our analysis shows significant overlap in incorrect queries across NL2SQL approaches, highlighting shared limitations. Many unresolved errors arise from inaccuracies in the BIRD dataset’s gold SQL annotations. Future work should focus on improving benchmark annotation quality to enable more reliable evaluations and meaningful progress.}

\begin{table}[t]
  \centering
  \caption{Effectiveness and efficiency on Query Revision on BIRD dev with DeepSeek-V3 vs GPT-4o-mini.
  We report \texttt{CI/I2C/E2C} and \#Tokens, \#LLM Calls. The best/second-best results are \fst{bold+underlined}/\snd{underlined} within each LLM.}
  \vspace{-0.4em}
  \scriptsize
  \renewcommand{\arraystretch}{1.0}
  \setlength\tabcolsep{1.2pt}
  \label{tab:revision_effectiveness_bird}
  \resizebox{\columnwidth}{!}{%
  \begin{tabular}{l|ccc|cc|ccc|cc}
    \toprule
    \multirow{2}{*}{\makecell{\\Approach}} &
      \multicolumn{5}{c|}{\textbf{DeepSeek-V3}} &
      \multicolumn{5}{c}{\textbf{GPT-4o-mini}} \\
    \cline{2-11}
      & \makecell{CI\\(\%)} & \makecell{I2C\\(\%)} & \makecell{E2C\\(\%)} & \makecell{\#Tok.} & \makecell{\#Calls}
      & \makecell{CI\\(\%)} & \makecell{I2C\\(\%)} & \makecell{E2C\\(\%)} & \makecell{\#Tok.} & \makecell{\#Calls} \\
    \midrule
    C3-SQL                 & -0.51 &  1.43 &  4.97 & --          & --   &  2.56 &  2.52 &  8.33 & --  & -- \\
    DIN-SQL                &  0.66 &  2.02 & 21.33 & 4948        & \fst{1}   & -0.11 &  4.16 & 19.12 & 4791  & \fst{1} \\
    MAC-SQL                &  3.55 &  3.66 & 13.16 & \fst{1566}  & \snd{2}   &  4.45 &  1.45 & 21.84 & \snd{1729}  & \snd{2}    \\
    CHESS\textsubscript{(IR,SS,CG)}
                           & \snd{6.11} &  6.89 & 29.36 & \snd{1844} & \fst{1}   &  6.72 &  10.49 & 35.29 & \fst{1683}  & \fst{1} \\
    CHESS\textsubscript{(IR,CG,UT)}
                           &  2.18 & \fst{15.16} & \fst{47.54} & 106126     & 21   & \snd{14.04} & \snd{16.09} & \snd{38.38} & 115055 & 23.19 \\
    GSR                    & \fst{7.27} & \snd{9.49} & \snd{38.10} & 2220       & \snd{2}    &  2.60 &  4.85 & 21.3 & 2449  & \snd{2}    \\
    E-SQL                  &  0.00 &  3.53 & 14.04 & 24924      & \snd{2}    &  1.91 &  7.52 & 28.29 & 25023 & \snd{2}    \\
    RSL-SQL                &  4.97 &  4.69 & 17.95 & 3315       & \snd{2}    &  9.15 & 10.06 & 35.86 & 2802  & \snd{2}    \\
    OpenSearch-SQL         &  3.16 &  8.67 & 34.04 & 6652       & 4.6  & \fst{14.07} & \fst{17.71} & \fst{51.20} & 12969 & 6.64 \\
    \bottomrule
  \end{tabular}%
  }
\end{table}

\begin{table}[t]
  \centering
  \caption{Effectiveness and efficiency on Query Revision on ScienceBenchmark dev with  DeepSeek-V3 vs GPT-4o-mini.
  We report \texttt{CI/I2C/E2C} and \#Tokens, \#LLM Calls. The best/second-best results are \fst{bold+underlined}/\snd{underlined}.}
  \vspace{-0.4em}
  \scriptsize
  \renewcommand{\arraystretch}{0.95}
  \setlength\tabcolsep{1.2pt}
  \label{tab:revision_effectiveness_sciencebenchmark}
  \resizebox{\columnwidth}{!}{%
  \begin{tabular}{l|ccc|cc|ccc|cc}
    \toprule
    \multirow{2}{*}{\makecell{\\Approach}} &
      \multicolumn{5}{c|}{\textbf{DeepSeek-V3}} &
      \multicolumn{5}{c}{\textbf{GPT-4o-mini}} \\
    \cline{2-11}
      & \makecell{CI\\(\%)} & \makecell{I2C\\(\%)} & \makecell{E2C\\(\%)} & \makecell{\#Tok.} & \makecell{\#Calls}
      & \makecell{CI\\(\%)} & \makecell{I2C\\(\%)} & \makecell{E2C\\(\%)} & \makecell{\#Tok.} & \makecell{\#Calls} \\
    \midrule
    C3-SQL & 0 & 0 & 0 & -- & --
           & 0 & 0 & 0 & -- & -- \\
    DIN-SQL & 1.27 & 2.99 & 14.29 & 4349 & \fst{1}
            & 2.34 & 1.38 & 15.38 & 4156 & \fst{1} \\
    MAC-SQL & 1.42 & 2.34 & 10 & 2395 & \snd{2}
            & 0.76 & 3.20 &  9.52 & 2676 & \snd{2} \\
    CHESS\textsubscript{(IR,SS,CG)}
            & \snd{19.47} & \fst{14.55} & 0 & \snd{2319} & \fst{1}
            & \fst{26.32} & \fst{13.30} & 14.29 & \snd{1683} & \fst{1} \\
    CHESS\textsubscript{(IR,CG,UT)}
            & \snd{1.83} & 9.32 & \fst{52.94} & 201372 & 23.42
            & \snd{20.49} & \snd{12.14} & \fst{43.24} & 115055 & 23.19 \\
    GSR     & -24.8 & \snd{14.08} & \snd{33.33} & \fst{1757} & \snd{2}
            & 4.96 & 5.13 & 18.18 & \fst{1498} & \snd{2} \\
    E-SQL   & -2.45 & 4.96 & 13.33 & 15088 & \snd{2}
            & 5.17 & 6.57 & 4.35 & 21360 & \snd{2} \\
    RSL-SQL & 1.47 & 0.00 & 4.44 & 6616 & \snd{2}
            & 5.79 & 0 & 35 & 3720 & \snd{2} \\
    OpenSearch-SQL
            & -1.74 & 7.41 & 3.88 & 18575 & 7.17
            & 6.35 & 7.14 & \snd{33.33} & 34752 & 11.33 \\
    \bottomrule
  \end{tabular}%
  }
\end{table}

\subsubsection{Detailed analysis on Query Revision performance.}
\label{sec:detailed_analysis_query_revision_performance}

We evaluate the Query Revision module using Correct-Improvement (CI), Incorrect-to-Correct (I2C), and Error-to-Correct (E2C), along with efficiency (\#tokens/\#LLM calls). The consolidated results for BIRD and ScienceBenchmark are reported in Table~\ref{tab:revision_effectiveness_bird} and Table~\ref{tab:revision_effectiveness_sciencebenchmark}, while the full details are in Appendix~\ref{appendix:revision_in_depth_analysis_full}.

\textbf{BIRD.} Using DeepSeek-V3, methods that employ light, execution-guided strategies, notably \textsc{GSR} and \textsc{CHESS}\textsubscript{(IR,SS,CG)}, obtain the highest CI with few calls and low tokens, while still posting healthy I2C/E2C. In contrast, unit-test-enabled pipelines such as \textsc{CHESS}\textsubscript{(IR,CG,UT)} achieve high I2C/E2C but only modest CI and at a very high cost. However, under GPT-4o-mini, the ranking shifts: breadth-oriented systems (\textsc{OpenSearch-SQL}, \textsc{CHESS}\textsubscript{(IR,CG,UT)}) show the best CI on BIRD, indicating that GPT-4o-mini responds better to selection or unit-test style prompts than DeepSeek-V3. Compact pipelines (\textsc{DIN-SQL}, \textsc{TA-SQL}, \textsc{CHESS}\textsubscript{(IR,SS,CG)}) remain cost-efficient with smaller but steady CI.

\textbf{ScienceBenchmark.} The results are polarized. \textsc{CHESS}\textsubscript{(IR,SS,CG)} delivers a large CI with a single call and low tokens. \textsc{CHESS}\textsubscript{(IR,CG,UT)} also posts high CI but at a very high budget. Several systems—even with decent I2C/E2C—show small or negative CI, meaning they flip correct queries to wrong during revision (i.e., high C2I risk).

\textbf{Error Analysis.} Across all four blocks, I2C/E2C improvements do not guarantee a high CI. When revision policies are aggressive, they may “fix” incorrect/error queries but also damage originally correct ones (C2I), netting a small or negative CI, as shown in the full detailed results in Appendix~\ref{appendix:revision_in_depth_analysis_full}. This explains the divergence we observe between good I2C/E2C figures and underwhelming CI on ScienceBenchmark/DeepSeek-V3 for several methods. Since our candidate generation analysis showed that IR (semantically misaligned yet executable SQL) dominates remaining errors, revision should prioritize semantic alignment and de-risk edits on already-correct SQL.

\textit{\textbf{Insight 8:} Future exploration should focus on strategies not only to correct wrong answers, but also to avoid altering originally correct queries into incorrect ones. Methods that incorporate robust "do-no-harm" safeguards for originally correct SQL can be beneficial.}

Note that even the highest CI, achieved in the Query Revision section, is just merely 26.32\%, which falls significantly short of a satisfactory threshold. Moreover, methods such as CHESS\textsubscript{(IR,CG,UT)} entail substantial token consumption and frequent LLM invocations, rendering them impractical for real-world deployment.

\textit{\textbf{Insight 9:} All of the current Query Revision strategies exhibit limited effectiveness in improving the Correct Rate, underscoring a crucial need to develop methods that are both genuinely effective—achieving significant accuracy improvements—and practically efficient, minimizing cost and latency.}

Additionally, for approaches that produce multiple queries, as shown in Figure \ref{fig:pass_at_k_revision}, while Correct Rates consistently improve with increasing \texttt{Pass@k}, the final Correct Rates after Query Revision remain notably lower, even below those at \texttt{Pass@5}. This indicates that the Query Revision module currently fails to approach the upper-bound effectiveness suggested by \texttt{Pass@5}.

\textit{\textbf{Insight 10:} The gap between \texttt{Pass@k} upper bounds and post-revision results identifies Query Revision as a key bottleneck, highlighting the need for more effective strategies to fully capitalise on the diversity of multiple candidates.}

\subsubsection{Practical guide.}
\label{sec:query_revision_practical_guide}

Our analysis reveals that current Query Revision strategies achieve limited effectiveness, with the predominant failure being the inability to correct queries that are syntactically valid but semantically misaligned.
To address this critical bottleneck, we recommend a dual-strategy revision approach that combines execution-guided feedback with conservative validation. First, implement execution-result-based semantic diagnosis that analyzes not just error messages but also the structure and content of returned results—comparing result schemas, row counts, and sample values against expected patterns derived from the natural language query.
Second, adopt a conservative revision policy that explicitly validates whether originally correct queries remain correct after revision.
Concretely, before committing a revision, execute both the original and revised queries and apply a consistency check: if the original query succeeded and the revised query produces different results, require additional validation steps (e.g., LLM-based semantic equivalence checking or unit test verification) before accepting the revision.

For systems employing multiple-candidate generation, our findings suggest that the query selection problem is as critical as the revision problem itself: the gap between Pass@5 upper bounds and post-revision performance indicates that current selection mechanisms fail to identify the best candidate reliably. We recommend implementing ensemble-based selection that combines multiple signals: execution success, result plausibility (e.g., non-empty results for queries expecting data), query complexity metrics, and LLM-based confidence scores.

Finally, for real-world deployment, given the substantial token costs of iterative revision (up to 100K+ tokens for multi-turn approaches), practitioners should implement early stopping mechanisms based on execution feedback to avoid unnecessary costs.

\subsection{End-to-End Overall Evaluation}
\label{sec:end_to_end_overall_evaluation}

\begin{table}[t]
\setlength\tabcolsep{1.5pt}
\renewcommand\arraystretch{1.0}
\small
\caption{End-to-end results on BIRD dev with two LLMs. We report \textit{Correct Rate}, efficiency (\#Tokens, \#LLM Calls) and \textit{Cost}. The best/second-best results are \fst{bold+underlined}/\snd{underlined} within each LLM block.}
\centering
\resizebox{\columnwidth}{!}{%
\begin{tabular}{l|rrrr|rrrr}
\toprule
\multirow{2}{*}{\makecell{\\Approach}}
& \multicolumn{4}{c|}{\textbf{DeepSeek-V3}}
& \multicolumn{4}{c}{\textbf{GPT-4o-mini}} \\
\cline{2-9}
& \makecell{Correct\\Rate (\%)} & \makecell{\#Tok-\\ens} & \makecell{\#LLM\\Calls} & \makecell{Cost\\(\$)}
& \makecell{Correct\\Rate (\%)} & \makecell{\#Tok-\\ens} & \makecell{\#LLM\\Calls} & \makecell{Cost\\(\$)} \\
\midrule
C3-SQL                 & 37.94 & 27673  & 40    & 0.0126 & 36.57 & 12332  & 40    & 0.0067 \\
DIN-SQL                & 59.97 & 18988  & \fst{3}    & 0.0038 & 56.58 & 18511  & \fst{3}    & 0.0024 \\
MAC-SQL                & 60.82 &  7574  & \snd{4}     & 0.0019 & 55.08 &  7867  & \snd{4}     & \snd{0.0013} \\
CHESS\textsubscript{(IR,SS,CG)} & 63.43 & 310756 & 80.6  & 0.0604 & 57.01 & 300061 & 80.5  & 0.0364 \\
CHESS\textsubscript{(IR,CG,UT)} & \snd{64.28} & 340601 & 41.8  & 0.0816 & \snd{62.82} & 343232 & 43.2  & 0.0525 \\
TA-SQL                 & 59.71 &  \snd{6842}  & \fst{3}    & \snd{0.0017} & 53.06 &  \snd{6546}  & \fst{3}    & \fst{0.0007} \\
GSR                    & 61.54 &  \fst{6066}  & \snd{4}     & \fst{0.0015} & 54.04 &  \fst{4455}  & \snd{4}     & \fst{0.0007} \\
E-SQL                  & 57.50 & 38903  & \fst{3}    & 0.0074 & 59.13 & 38054  & \fst{3}    & 0.0047 \\
RSL-SQL                & 61.93 & 20475  & 8     & 0.0041 & 60.63 & 10549  & 8     & 0.0013 \\
OpenSearch-SQL         & \fst{68.19} & 113446 & 28.6  & 0.0242 & \fst{64.47} & 28569  & 10.7  & 0.0057 \\
\bottomrule
\end{tabular}
}
\label{tab:overall_analysis_details_bird}
\vspace{-1em}
\end{table}

\begin{table}[t]
\setlength\tabcolsep{1.5pt}
\renewcommand\arraystretch{1.0}
\small
\caption{End-to-end results on ScienceBenchmark dev with two LLMs. We report \textit{Correct Rate}, efficiency (\#Tokens, \#LLM Calls) and \textit{Cost}. The best/second-best results are \fst{bold+underlined}/\snd{underlined} within each LLM block.}
\centering
\resizebox{\columnwidth}{!}{%
\begin{tabular}{l|rrrr|rrrr}
\toprule
\multirow{2}{*}{\makecell{\\Approach}}
& \multicolumn{4}{c|}{\textbf{DeepSeek-V3}}
& \multicolumn{4}{c}{\textbf{GPT-4o-mini}} \\
\cline{2-9}
& \makecell{Correct\\Rate (\%)} & \makecell{\#Tok-\\ens} & \makecell{\#LLM\\Calls} & \makecell{Cost\\(\$)}
& \makecell{Correct\\Rate (\%)} & \makecell{\#Tok-\\ens} & \makecell{\#LLM\\Calls} & \makecell{Cost\\(\$)} \\
\midrule
C3-SQL                 & 54.18 & 30979   & 40        & 0.0152  & \snd{50.84} & 14179   & 40        & 0.0078 \\
DIN-SQL                & 53.51 & 17299   & \fst{3}   & 0.0035  & 43.81       & 16660   & \fst{3}        & 0.0022 \\
MAC-SQL                & 47.83 &  9290 & \snd{4}     & 0.0023  & 44.48       &  9577 & \snd{4}          & 0.0015 \\
CHESS\textsubscript{(IR,SS,CG)} & 45.30 & 355983  & 92.59     & 0.0677  & \fst{57.01} & 300061  & 80.5       & 0.0364 \\
CHESS\textsubscript{(IR,CG,UT)} & \snd{55.85} & 460941  & 43.42     & 0.1018  & 49.16       & 393208  & 47.6       & 0.0428 \\
TA-SQL                 & \fst{59.53} & \snd{7867} & \fst{3}   & \snd{0.0018} & 50.17       & \snd{7572} & \fst{3}    & \snd{0.0008} \\
GSR                    & 30.43 & \fst{2581} & \snd{4}       & \fst{0.0004} & 42.47       & \fst{2264} & \snd{4}          & \fst{0.0004} \\
E-SQL                  & 53.18 & 21630   & \fst{3}   & 0.0044  & 40.80       & 34096   & \fst{3}        & 0.0042 \\
RSL-SQL                & 46.15 & 10546   & 8        & 0.0021  & 42.81       & 11990   & 8          & 0.0018 \\
OpenSearch-SQL         & 37.79 & 128204  & 31.08     & 0.0272  & 44.82       & 49371   & 15.34       & 0.0084 \\
\bottomrule
\end{tabular}
}
\label{tab:overall_analysis_details_sciencebenchmark}
\vspace{-1em}
\end{table}

We summarize the end-to-end overall evaluation of all approaches in Table \ref{tab:overall_analysis_details_bird} and Table \ref{tab:overall_analysis_details_sciencebenchmark}. The cost per task is estimated based on the pricing guidelines of DeepSeek-V3\footnote{https://api-docs.deepseek.com/} and GPT-4o-mini\footnote{https://platform.openai.com/docs/pricing}. To simulate a realistic pricing scenario, we assume that half of the prompt tokens result in cache hits and the other half in cache misses.

\textbf{BIRD.} With DeepSeek-V3, \textsc{OpenSearch-SQL} attains the highest CR (68\%),
about 29 calls and 113K tokens per query. CHESS variants are similarly accurate yet even more expensive. In contrast, compact pipelines—\textsc{TA-SQL}, \textsc{MAC-SQL}, \textsc{GSR}—reach 59–61\% CR with 1–4 calls and 6–8K tokens, yielding much lower cost. The results obtained using GPT-4o-mini are highly similar, indicating the performance of the evaluated approaches is stable across LLMs.

\textbf{ScienceBenchmark.} With DeepSeek-V3, leadership shifts to \textsc{TA-SQL} (60\% CR) using only 3 calls and 8K tokens. Breadth-heavy systems (CHESS\textsubscript{(IR,CG,UT)}, OpenSearch-SQL) consume 40–90+ calls and 300K–460K tokens yet do not dominate CR. With GPT-4o-mini, \textsc{CHESS}\textsubscript{(IR,SS,CG)} achieves the best CR (57\%) but at a very high cost (80 calls and 300K tokens). \textsc{TA-SQL} retains strong cost efficiency. The ranking shift indicates there is no universal winner on the ScienceBenchmark dataset: the best method depends on both the dataset and the backbone LLM.

\textit{\textbf{Insight 11:} End-to-end rankings are LLM and dataset-dependent; validate on the target database(s) with the intended LLM before adoption is necessary.}

\textbf{Cost-efficiency.}
While OpenSearch-SQL, CHESS\textsubscript{(IR,CG,UT)}, and CHESS\textsubscript{(IR,SS,CG)} achieve superior accuracy on some workloads, they incur substantial token costs and LLM invocations. Recent studies, such as CHASE-SQL~\cite{pourreza2025chasesql} and MCS-SQL~\cite{lee2024mcs}, have also shown outstanding performance by increasing LLM calls to generate a large pool of candidate queries.
However, we believe that while increasing the number of LLM calls can improve accuracy, this approach may lack adaptability in real-world deployment.

\textit{\textbf{Insight 12:} While scaling LLM calls and token usage can enhance accuracy on some datasets, such strategies may undermine scalability and cost-effectiveness. Future work should aim to achieve high accuracy without proportional growth in token consumption.}

\subsubsection{Practical guide.}
\label{sec:end_to_end_practical_guide}

We recommend reporting CR@\emph{budget} (e.g., CR at a fixed call or token cap) and selecting methods along the Pareto frontier instead of by CR alone. To achieve high accuracy without huge token consumption, we recommend implementing adaptive module selection that dynamically adjusts computational strategies based on query characteristics and system confidence signals. Specifically, design a query complexity classifier (which can be a lightweight model or rule-based heuristic) that categorizes incoming queries into simple, moderate, and challenging classes based on features such as question length, number of entities mentioned, presence of aggregations or joins, and schema size. For simple queries, employ streamlined pipelines that skip expensive multi-stage schema pruning and multi-candidate generation, directly applying single-pass methods
For moderate queries, selectively enable key optimization strategies—such as two-stage schema pruning and single-turn execution-guided revision—that hopefully provide acceptable accuracy at reasonable cost.
Reserve expensive strategies like multi-candidate generation with consistency-based selection and iterative multi-turn revision exclusively for challenging queries.
Additionally, implement confidence-based early termination: if a candidate query executes successfully, returns non-empty results with the expected schema structure, and receives high LLM confidence scores, terminate the pipeline early to avoid unnecessary revision costs.
This adaptive approach can potentially reduce average token consumption while maintaining the accuracy of always-expensive strategies, thereby enabling practical deployment at scale.

\subsection{New Discoveries on Existing Datasets}
\label{sec:datasets_analysis}

We conducted an in-depth analysis of BIRD and identified three key limitations that compromise its reliability as follows. Detailed examples are provided in Appendix~\ref{appendix:incorrect_gold_sql_example}.

\subsubsection{Inaccurate Gold SQL Queries.}
\label{sec:inaccurate_gold_sql}

As shown in Section~\ref{sec:result_query_revision}, some NL queries in BIRD are paired with the wrong Gold SQL queries. Although~\cite{liu2025nl2sql-bugs} recently identified 106 incorrectly annotated queries in the BIRD dataset, their results still have not exhaustively covered all annotation errors. The prevalence of such annotation errors raises concerns about the reliability and validity of existing datasets.

\subsubsection{Strict Evaluation Rules.}

The BIRD benchmark currently assesses each NL question against a single gold SQL query, an assumption that proves overly restrictive in many realistic scenarios~\cite{FloratouPZDHTCA24}. This over-strictness calls for more flexible and semantically aware evaluation rules—such as result-set equivalence or graded relevance metrics—to foster fairer assessments of future NL2SQL solutions.

\subsubsection{Semantic Ambiguity.}

Existing datasets inevitably contain questions that exhibit a certain degree of semantic ambiguity~\cite{FloratouPZDHTCA24}. Mitigating the effect of such ambiguity will require either (i) rewriting questions to eliminate ambiguity or (ii) devising evaluation rules that recognize ambiguity and reward semantically equivalent answers. Developing formal methods to define, detect, and quantify semantic ambiguity remains an open problem and constitutes a promising avenue for future research.

\textit{\textbf{Insight 13:} Systematic auditing of gold annotations, adoption of evaluation rules that acknowledge multiple semantically equivalent queries, and explicit modeling of semantic ambiguity are crucial for reliable assessment and methodological progress. Tackling these issues, notably present in the BIRD dataset, will enable more accurate evaluations and advance robust NL2SQL solutions.}

\subsubsection{Practical guide.}
\label{sec:dataset_practical_guide}

we provide the following recommendations for practitioners and benchmark developers.

First, to mitigate the impact of inaccurate gold annotations, we recommend implementing multi-reference evaluation where generated queries are validated against multiple acceptable SQL variants rather than a single gold query. For production systems, establish a human-in-the-loop validation process for queries that execute successfully but are marked incorrect by automated evaluation, as these cases often indicate annotation errors rather than system failures.

Second, to address strict evaluation rules that fail to recognize semantically equivalent results, adopt flexible evaluation metrics such as partial credit scoring or schema-normalized comparison. When deploying NL2SQL systems, provide users with query explanation mechanisms that clarify what information the generated SQL retrieves, enabling users to verify semantic correctness even when result formatting differs from expectations.

Third, to handle semantic ambiguity in natural language questions, implement ambiguity detection and clarification mechanisms: use LLM-based analysis to identify potentially ambiguous terms, generate clarification questions for users, or provide multiple query interpretations with explanations. For benchmark developers, we strongly advocate for systematic quality assurance protocols including: (1) multi-annotator consensus requirements for gold SQL creation, (2) automated consistency checking to detect annotation errors, and (3) explicit ambiguity annotations marking questions with multiple valid interpretations.

Furthermore, we recommend developing semantically-aware evaluation frameworks that can recognize and credit semantically equivalent but syntactically different SQL queries, moving beyond simple string matching or single-reference execution accuracy. These quality improvements are essential not only for fair benchmark evaluation but also for training more robust NL2SQL systems, as models trained or evaluated on noisy benchmarks may learn to replicate annotation errors rather than develop genuine semantic understanding. More details are provided in Appendix~\ref{appendix:incorrect_gold_sql_example}.

\subsection{Leveraging {\toolname} for NL2SQL System Development}
\label{usability}

The modular architecture of {\toolname} enables practitioners to adopt a \textbf{compositional optimization} approach to NL2SQL development, analogous to assembling building blocks where each module can be independently diagnosed, optimized, and composed.
To leverage this capability effectively, we recommend the following systematic workflow.
First, establish a performance profile by running {\toolname} with different baselines on your target dataset to generate a comprehensive performance profile across all three core modules. This profile identifies the primary bottleneck.
Second, apply targeted optimizations to the bottleneck module while monitoring cross-module impacts.
Third, conduct a cost-benefit analysis at the module level using our effectiveness and efficiency metrics.
Finally, leverage {\toolname} for A/B testing and continuous improvement.
%
By treating NL2SQL as a modular optimization problem where each component can be independently analyzed, improved, and composed, rather than a monolithic system design challenge, practitioners can systematically navigate the accuracy-efficiency trade-off space and construct systems tailored to their specific deployment requirements. We provide a case study of leveraging {\toolname}
for NL2SQL system development in Appendix~\ref{appendix:case_study}.




%% file: sections/related-work.tex
\section{Related Work}
\label{sec:relatedwork}

\emph{LLM-based NL2SQL.}
Since LLMs have demonstrated distinctive emergent abilities~\cite{wei2022emergent}, LLM-based NL2SQL methods have become the most prominent solutions in the current NL2SQL landscape, as described in recent surveys \cite{DBLP:journals/tkde/LiuSLMJZFLTL25, zhu2024largelanguagemodelenhanced, shi2024surveyemployinglarge, hong2024next, mohammadjafari2024naturallanguagesqlreview}. Detailed discussions are presented in Appendix~\ref{appendix:related_work}.

\emph{NL2SQL Datasets and Dataset Evaluation.}
Several datasets have been proposed recently to facilitate the development and evaluation of NL2SQL solutions.
WikiSQL~\cite{dataset-wikisql} and Spider~\cite{dataset-spider} published several cross-domain datasets, which mainly contain lightweight databases and focus on database schema.
BIRD~\cite{dataset-bird} focuses on large databases and external knowledge, and has become a widely used NL2SQL dataset nowadays.
ScienceBenchmark~\cite{dataset-sciencebenchmark} is a highly domain-specific NL2SQL dataset in the field of scientific data analysis, which contains real-world complex queries and large databases.
Recently proposed Spider-2.0~\cite{lei2025spider2} provides a more realistic enterprise-level NL2SQL benchmark, encompassing multiple database systems, diverse SQL dialects, and numerous challenging tasks from real data engineering pipelines.
In the line of related work that evaluates NL2SQL datasets, Mitsopoulou and Koutrika proposed a comprehensive analysis of existing NL2SQL datasets, and provided valuable insights into their capabilities and limitations, and how they affect training and evaluation of NL2SQL systems \cite{mitsopoulou2025analysis}.
\citet{liu2025nl2sql-bugs} further developed a benchmark for detecting and categorizing semantic errors in NL2SQL. ~\citet{SQLDriller} presents a method for detecting and fixing the errors in NL2SQL translation.

\emph{NL2SQL System Evaluations.}
Several experimental studies have been proposed to evaluate NL2SQL solutions. For example, ~\citet{dataset-dr-spider} evaluated the robustness of different NL2SQL models. ~\citet{pourreza-rafiei-2023-evaluating} studied the limitations of existing evaluation metrics and conducted a benchmark through human evaluation and query rewriting. ~\citet{gao2023text} undertook an assessment examining multiple prompting strategies and fine-tuning methods. ~\citet{zhang2024benchmarking} evaluated the performance of different LLMs in sub-tasks of NL2SQL pipeline. ~\citet{nl2sql360} proposed a testbed for evaluating NL2SQL systems from different perspectives.

Unlike previous efforts, our work, \texttt{\toolname}, is the first to introduce a set of fine-grained evaluation metrics for each module of the NL2SQL pipeline, develop a benchmarking framework, and conduct a comprehensive benchmarking of diverse strategies at the modular level. We summarize the differences between \texttt{\toolname} and existing NL2SQL evaluation frameworks in Table \ref{tab:related_work_comparison}.

\begin{table}[h]
  \caption{Comparison of evaluation frameworks for NL2SQL datasets and systems.}
  \label{tab:related_work_comparison}
  \setlength\tabcolsep{0.5pt}
  \resizebox{\columnwidth}{!}{%
    \begin{tabular}{l|c|c|c|c|c|c}
      \hline
      Benchmarking Framework & \makecell{Evaluation\\Objectives} & \makecell{E2E\\ Eval} & \makecell{Modular\\ Eval} & \makecell{Fine-grain-\\ed Metrics} & \makecell{Error \\ Analysis} & \makecell{Practical \\ Guides} \\
      \hline
      Text2SQL Benchmarks \cite{mitsopoulou2025analysis} & Datasets & -- & -- & -- & \textbf{Yes} & -- \\
      \hline
      SQLDriller \cite{SQLDriller} & Datasets & -- & -- & -- & \textbf{Yes} & -- \\
      \hline
      NL2SQL-BUGs \cite{liu2025nl2sql-bugs} & Datasets & -- & -- & -- & \textbf{Yes} & -- \\
      \hline
      DR-Spider \cite{dataset-dr-spider} & Systems & \textbf{Yes} & No & No & \textbf{Yes} & Partial \\
      \hline
      Cross-Domain Text2SQL \cite{pourreza-rafiei-2023-evaluating} & Systems & \textbf{Yes} & No & No & \textbf{Yes} & Partial \\
      \hline
      Text2SQL by LLMs \cite{gao2023text} & Systems & \textbf{Yes} & No & No & No & Partial \\
      \hline
      Benchmarking Text2SQL \cite{zhang2024benchmarking} & Systems & \textbf{Yes} & No & No & \textbf{Yes} & Partial \\
      \hline
      NL2SQL360 \cite{nl2sql360} & Systems & \textbf{Yes} & No & No & No & Partial \\
      \hline
      {\toolname}~(ours) & Systems & \textbf{Yes} & \textbf{Yes} & \textbf{Yes} & \textbf{Yes} & \textbf{Detailed} \\
      \hline
    \end{tabular}
  }
\end{table}


%% file: sections/discussion.tex
\section{Limitations}

\emph{Limited Scope of Evaluated Approaches.}
Our evaluation encompasses a selection of representative NL2SQL approaches from the BIRD leaderboard.
However, due to code availability constraints and limited computational resources, not all existing NL2SQL methods were included.
For instance, the studies~\cite{li2024codes, pourreza2025chasesql, li2025omnisql, xiyansql, cohere2025command, pourreza2025reasoning, sheng2025csc} that fine-tune specialized models for specific tasks within the NL2SQL pipeline were excluded from our benchmarking. Future work should incorporate a broader array of approaches to enable a more comprehensive evaluation of the current landscape of NL2SQL systems.

\emph{Lack of Industry-Level Evaluation.}
Current benchmarks may not fully reflect the complexities of real-world industry workloads, including schema intricacy and query ambiguity. Future research should incorporate more realistic, industry-level evaluations to better capture practical deployment challenges.



%% file: sections/conclusion.tex
\section{Conclusions}
We have systematically addressed the critical need for a unified, modular evaluation of LLM-enabled NL2SQL approaches.
Specifically, we conducted a comprehensive review of the three core modules
\emph{Schema Selection}, \emph{Candidate Generation}, and \emph{Query Revision}.
We proposed a novel set of fine-grained metrics and developed {\toolname}, a modular, multi-agent benchmarking framework.
Our evaluation of representative NL2SQL approaches
highlights substantial opportunities for improvement in both accuracy and computational efficiency
Our in-depth analysis exposed critical shortcomings in current benchmark datasets and evaluation rules, underscoring the urgency for developing more accurate, robust, and standardized evaluation resources.
By establishing this foundational benchmarking framework, our work aims to provide useful insights and practical guides for future NL2SQL development, ultimately benefiting both academic research and enterprise applications.

%% file: sections/appendix.tex
\clearpage
\appendix

\section{Appendix}

\subsection{JSON Structure Used to Collect Results from NL2SQL Solutions}
\label{appendix:json_structure_result_collection}

We present an example of the JSON format used by our Result Collecting Agent to collect the results of each NL2SQL solution as follows:

\begin{lstlisting}[language=json]
{{
 "node_type": "schema_selection",
 "question": "How many cards are there with toughness of 99?",
 "extracted_schema": { "cards": ["id", "toughness"] }
 "token_cost": 1200,
 "llm_calls": 1
 },
 {
 "node_type": "candidate_generation",
 "question": "How many cards are there with toughness of 99?",
 "SQL": "SELECT COUNT(id) FROM cards WHERE toughness = 99",
 "token_cost": 1200,
 "llm_calls": 1
 },
 {
 "node_type": "query_revision",
 "question": "How many cards are there with toughness of 99?",
 "SQL": "SELECT COUNT(id) FROM cards WHERE toughness = 99",
 "token_cost": 1200,
 "llm_calls": 1
 }}
\end{lstlisting}

The meanings of JSON fields in the collected files are:

\begin{itemize}[nosep, noitemsep]
  \item \texttt{node\_type}: the module that produced this record.
  \item \texttt{question}: the original natural language query \(Q\).
  \item \texttt{extracted\_schema}: a mapping of selected tables and columns.
  \item \texttt{SQL}: the SQL generated by the model.
  \item \texttt{token\_cost}: the token cost for this module.
  \item \texttt{llm\_calls}: the number of of LLMs invocations.
\end{itemize}

\subsection{Additional Results for Schema Selection Module}
\label{appendix:schema_selection}
In this section, we present the results for the Schema Selection module obtained using GPT-4o-mini on the BIRD dataset, using DeepSeek-V3 on the ScienceBenchmark dataset, and using GPT-4o-mini on the ScienceBenchmark dataset, as shown in Table \ref{tab:schema_selection_bird_gpt4omini}, Table \ref{tab:schema_selection_bench_deepseek}, and Table \ref{tab:schema_selection_bench_gpt4omini}, respectively.

\begin{table}[h]
  \centering
  \caption{Analysis results of Schema Selection on BIRD dev set using GPT-4o-mini. The best results are in \fst{bold and underlined} while the second-best results are \snd{underlined}}.
  \vspace{-0.3em}
  \setlength\tabcolsep{1pt}
  \small
  \renewcommand{\arraystretch}{1.0}
  \label{tab:schema_selection_bird_gpt4omini}
  \resizebox{\columnwidth}{!}{%
  \begin{tabular}{l|c|c|c|c|c|c|c|c}
     \toprule
        \multirow{2}{*}{\makecell{\\Approach}}   & \multicolumn{3}{c|}{Table Selection}  & \multicolumn{3}{c|}{Column Selection}  & \multicolumn{2}{c}{Efficiency} \\ \cline{2-9}
        &  \makecell{Precis-\\ion (\%)}  &  \makecell{Recall\\(\%)} &  \makecell{F1-score\\(\%)} &  \makecell{Precis-\\ion (\%)} &  \makecell{Recall\\(\%)} &  \makecell{F1-score\\(\%)} &  \makecell{\#Tok-\\ens} &  \makecell{\#LLM\\calls}\\
        \midrule
        C3-SQL                & 49.79  & 98.46  & 64.35  & 26.80  & 93.75  & 40.30  & 10979 & 20   \\
        DIN-SQL               & \snd{92.08}  & 95.89  & \snd{92.83}  & \snd{87.49}  & 85.51  & \fst{85.47}  & 7197   & \fst{1}   \\
        MAC-SQL               & 32.74  & \fst{99.84}  & 46.70  & 14.83  & \snd{96.63}  & 24.49  & \fst{3107}   & \fst{1}   \\
        CHESS\textsubscript{(IR,SS,CG)} & 90.96  & 95.80  & 92.17  & \fst{88.74}  & 74.14  & 78.96  & 297443   & 78.53   \\
        TA-SQL                & \fst{92.56}  & 95.38  & \fst{92.88}  & 81.78  & 88.44  & \snd{83.25}  & \snd{4119}   & \fst{1}   \\
        RSL-SQL               & 81.64  & 97.18  & 86.57  & 46.58  & 80.95  & 52.39  & 5528   & \fst{1}  \\
        OpenSearch-SQL        & 32.76  & \snd{99.49}  & 46.66  & 16.88  & \fst{97.14}  & 27.34  & 6616   & \snd{3}   \\
     \bottomrule
  \end{tabular}%
  }
\end{table}

\begin{table}[h]
  \centering
  \caption{Analysis results of Schema Selection on ScienceBenchmark dev set using DeepSeek-V3. The best results are in \fst{bold and underlined} while the second-best results are \snd{underlined}}.
  \vspace{-0.3em}
  \setlength\tabcolsep{1pt}
  \small
  \renewcommand{\arraystretch}{1.0}
  \label{tab:schema_selection_bench_deepseek}
  \resizebox{\columnwidth}{!}{%
  \begin{tabular}{l|c|c|c|c|c|c|c|c}
     \toprule
        \multirow{2}{*}{\makecell{\\Approach}}   & \multicolumn{3}{c|}{Table Selection}  & \multicolumn{3}{c|}{Column Selection}  & \multicolumn{2}{c}{Efficiency} \\ \cline{2-9}
        &  \makecell{Precis-\\ion (\%)}  &  \makecell{Recall\\(\%)} &  \makecell{F1-score\\(\%)} &  \makecell{Precis-\\ion (\%)} &  \makecell{Recall\\(\%)} &  \makecell{F1-score\\(\%)} &  \makecell{\#Tok-\\ens} &  \makecell{\#LLM\\calls}\\
        \midrule
        C3-SQL                & 49.50  & \snd{97.06}  & 62.98  & 30.69  & \snd{94.30}  & 43.64  & 18869 & 20   \\
        DIN-SQL               & 89.58  & 81.15  & 83.48  & \fst{79.80}  & 73.20  & \snd{73.92}  & 6802   & \fst{1}   \\
        MAC-SQL               & 18.56  & \fst{100}  & 28.93  & 11.15  & \fst{96.39}  & 18.42  & \fst{3380}   & \fst{1}   \\
        CHESS\textsubscript{(IR,SS,CG)} & \snd{91.23}  & 82.44  & 84.86  & \snd{76.48}  & 73.51  & 72.42  & 352686   & 92.59   \\
        TA-SQL                & \fst{92.72}  & 90.20  & \fst{90.13}  & 74.16  & 85.27  & \fst{76.39}  & \snd{5155}   & \fst{1}   \\
        RSL-SQL               & 86.53  & 91.76  & \snd{87.01}  & 40.59  & 88.17  & 50.39  & 7669   & \fst{1}  \\
        OpenSearch-SQL        & 12.35  & 67.89  & 19.28  & 7.91   & 66.79  & 13.44  & 6237   & \snd{3}   \\
     \bottomrule
  \end{tabular}%
  }
\end{table}

\begin{table}[h]
  \centering
  \caption{Analysis results of Schema Selection on ScienceBenchmark dev set using GPT-4o-mini. The best results are in \fst{bold and underlined} while the second-best results are \snd{underlined}}.
  \vspace{-0.3em}
  \setlength\tabcolsep{1pt}
  \small
  \renewcommand{\arraystretch}{1.0}
  \label{tab:schema_selection_bench_gpt4omini}
  \resizebox{\columnwidth}{!}{%
  \begin{tabular}{l|c|c|c|c|c|c|c|c}
     \toprule
        \multirow{2}{*}{\makecell{\\Approach}}   & \multicolumn{3}{c|}{Table Selection}  & \multicolumn{3}{c|}{Column Selection}  & \multicolumn{2}{c}{Efficiency} \\ \cline{2-9}
        &  \makecell{Precis-\\ion (\%)}  &  \makecell{Recall\\(\%)} &  \makecell{F1-score\\(\%)} &  \makecell{Precis-\\ion (\%)} &  \makecell{Recall\\(\%)} &  \makecell{F1-score\\(\%)} &  \makecell{\#Tok-\\ens} &  \makecell{\#LLM\\calls}\\
        \midrule
        C3-SQL                & 49.25  & 95.39  & 62.19  & 29.17  & 91.50  & 41.51  & 12829 & 20   \\
        DIN-SQL               & \snd{90.18}  & 81.71  & 83.65  & \snd{77.23}  & 68.77  & \snd{70.65}  & 6571   & \fst{1}   \\
        MAC-SQL               & 18.56  & \fst{100}  & 28.93  & 10.48  & \snd{96.22}  & 17.26  & \fst{3167}   & \fst{1}   \\
        CHESS\textsubscript{(IR,SS,CG)} & \fst{90.96}  & \snd{95.80}  & \fst{92.17}  & \fst{88.74}  & 74.14  & \fst{78.96}  & 341059   & 92.59   \\
        TA-SQL                & 90.17  & 86.36  & \snd{86.38}  & 67.98  & 79.63  & 70.19  & \snd{5049}   & \fst{1}   \\
        RSL-SQL               & 74.40  & 85.65  & 76.88  & 39.10  & 70.97  & 42.73  & 7453   & \fst{1}  \\
        OpenSearch-SQL        & 18.56  & \fst{100}  & 28.93  & 12.30  & \fst{97.90}  & 20.77  & 6264   & \snd{3}   \\
     \bottomrule
  \end{tabular}%
  }
\end{table}

\subsection{Additional Results for Candidate Generation Module}
\subsubsection{Error Analysis on Candidate Generation}
\label{appendix:error_analysis_candidate_generation}
We break execution errors (ER) into five categories: No Table/Column, No Function, Syntax Error, Timeout, and Others. Although our earlier analysis shows that incorrect-but-executable (IR) dominates overall failures, understanding the composition of ER is important for robustness and cost. The detailed results are shown in Table \ref{tab:candidate_generation_error_bird_ds}, Table \ref{tab:candidate_generation_error_bird_gpt4omini}, Table \ref{tab:candidate_generation_error_scib_deepseekv3} and Table \ref{tab:candidate_generation_error_scib_gpt4omini}.

\textbf{BIRD.} With DeepSeek-V3, OpenSearch-SQL exhibits the lowest total ER, with small contributions from all buckets; CHESS\textsubscript{(IR,SS,CG)} is also strong but sees a slightly higher Timeout share. With GPT-4o-mini, the lowest ER shifts to CHESS\textsubscript{(IR,SS,CG)}, while several methods show noticeably higher No Table/Column errors than with DeepSeek-V3.
Overall, on BIRD, execution failures are already rare for the top systems; the main movable pieces are schema-linking lapses (No Table/Column) and light syntax/dialect issues.

\textbf{ScienceBenchmark.} With DeepSeek-V3, patterns diverge: TA-SQL keeps ER very low, while OpenSearch-SQL shows a dramatic spike in Others. With GPT-4o-mini, CHESS\textsubscript{(IR,SS,CG)} achieves the lowest ER, while several methods suffer increased No Table/Column rates.
Overall, on denser, more realistic schemas, resource and plan-related failures (Timeout/Others) become salient for breadth-heavy generation, whereas IR/SS-style pipelines remain execution-stable.

\textbf{Backbone LLM effects.} Switching from DeepSeek-V3 to GPT-4o-mini increases No Table/Column errors for multiple methods on both datasets, suggesting backbone-specific schema-linking calibration matters. By contrast, Syntax Error remains low across all settings, indicating that most generators respect SQL grammar.

\begin{table}
  \centering
  \caption{Error analysis on BIRD dev set with DeepSeek-V3 for Candidate Generation.
  The best results are in \fst{bold and underlined} while the second-best results are \snd{underlined}.
  We use \texttt{Pass@1} results for approaches producing multiple candidates.}
  \setlength\tabcolsep{2pt}
  \renewcommand{\arraystretch}{1.0}
  \label{tab:candidate_generation_error_bird_ds}
  \small
  \resizebox{0.99\columnwidth}{!}{%
  \begin{tabular}{l|c|c|c|c|c|c}
     \toprule
    \multirow{2}{*}{\makecell{Approach}}
      & \multicolumn{5}{c|}{Error Rates (\%)}
      & \multirow{2}{*}{\makecell{Total\\(\%)}} \\
    \cmidrule(lr){2-6}
      & \makecell{No Tabl/Col}
      & \makecell{No Func}
      & \makecell{Syntax Err}
      & Timeout
      & Others
      &  \\
    \midrule
    C3-SQL                & 3.32  & \fst{0}      & \fst{0.65}  & 0.52  & 7.30  & 11.80 \\

    DIN-SQL               & 0.98  & 0.13  & 3.12  & 0.26  & \snd{0.39}  & 4.89 \\

    MAC-SQL               & 1.11  & \snd{0.07}  & 1.83  & 0.26  & 4.17  & 7.43 \\

    CHESS\textsubscript{(IR,SS,CG)} & \snd{0.46}  & 1.43  & 1.69  & 3.32  & \fst{0.20}  & 7.11 \\

    CHESS\textsubscript{(IR,CG,UT)} & 1.30  & \fst{0}      & 2.28  & \fst{0.20}  & \fst{0.20}  & \snd{3.98} \\

    TA-SQL                & 1.30  & 1.69  & 2.54  & \snd{0.13}  & \fst{0.20}  & 5.87 \\

    GSR                   & 2.41  & 0.58  & 1.63  & 0.52  & 0     & 5.48 \\

    E-SQL                 & 0.65 & 0.13 & 2.87 & 0.26 & 3.52 & 7.43 \\

    RSL-SQL               & \fst{0.39}  & 0.91  & 1.37  & 0.33  & 4.62  & 7.62 \\

    OpenSearch-SQL        & 0.78  & \snd{0.07}  & \snd{0.84}  & \fst{0.07}  & 1.30  & \fst{3.06} \\
    \bottomrule
  \end{tabular}%
  }
\end{table}

\begin{table}
  \centering
  \caption{Error analysis on \textbf{BIRD Dev using GPT-4o-mini for Candidate Generation}.
  The best results are in \fst{bold and underlined} while the second-best results are \snd{underlined}.
  We use \texttt{Pass@1} results for approaches producing multiple candidates.}
  \setlength\tabcolsep{2pt}
  \renewcommand{\arraystretch}{1.0}
  \label{tab:candidate_generation_error_bird_gpt4omini}
  \small
  \resizebox{0.99\columnwidth}{!}{%
  \begin{tabular}{l|c|c|c|c|c|c}
     \toprule
    \multirow{2}{*}{\makecell{Approach}}
      & \multicolumn{5}{c|}{Error Rates (\%)}
      & \multirow{2}{*}{\makecell{Total\\(\%)}} \\
    \cmidrule(lr){2-6}
      & \makecell{No Tabl/Col}
      & \makecell{No Func}
      & \makecell{Syntax Err}
      & Timeout
      & Others
      &  \\
    \midrule
    C3-SQL                      & 9.19 & \fst{0} & 1.96 & \snd{0.07} & 1.3  & 12.52 \\
    DIN-SQL                     & 5.48 & 0.46    & 2.41 & 0.2        & 0.33 & 8.87  \\
    MAC-SQL                     & 6.26 & 1.43    & 0.85 & 0.33       & 2.48 & 11.34 \\
    CHESS\textsubscript{(IR,SS,CG)} & \fst{2.54} & 0.72 & 1.11 & 0.91 & 0.26 & \fst{5.54} \\
    CHESS\textsubscript{(IR,CG,UT)} & 9.32 & 1.76 & 1.63 & \fst{0}    & \snd{0.2} & 12.91 \\
    TA-SQL                      & 7.37 & 4.82    & 0.78 & 0.26       & \fst{0.13} & 13.36 \\
    GSR                         & \fst{2.54} & \snd{0.39} & 1.43 & \snd{0.07} & 2.61 & \snd{7.04} \\
    E-SQL                       & 5.48 & 0.72    & 1.17 & \snd{0.07} & 2.48 & 9.91  \\
    RSL-SQL                     & \snd{4.24} & 1.5 & \snd{0.72} & 0.13 & 2.87 & 9.45 \\
    OpenSearch-SQL              & 6.13 & 0.85    & \fst{0.46} & 0.2 & 0.52 & 8.15 \\
    \bottomrule
  \end{tabular}%
  }
\end{table}

\begin{table}
  \centering
  \caption{Error analysis on \textbf{ScienceBenchmark Dev using DeepSeek-V3} for Candidate Generation.
  The best results are in \fst{bold and underlined} while the second-best results are \snd{underlined}.
  We use \texttt{Pass@1} results for approaches producing multiple candidates.}
  \setlength\tabcolsep{2pt}
  \renewcommand{\arraystretch}{1.0}
  \label{tab:candidate_generation_error_scib_deepseekv3}
  \small
  \resizebox{0.99\columnwidth}{!}{%
  \begin{tabular}{l|c|c|c|c|c|c}
     \toprule
    \multirow{2}{*}{\makecell{Approach}}
      & \multicolumn{5}{c|}{Error Rates (\%)}
      & \multirow{2}{*}{\makecell{Total\\(\%)}} \\
    \cmidrule(lr){2-6}
      & \makecell{No Tabl/Col}
      & \makecell{No Func}
      & \makecell{Syntax Err}
      & Timeout
      & Others
      &  \\
    \midrule
    C3-SQL                      & 3.01 & \fst{0} & \fst{0}    & \fst{0} & \fst{0}    & 3.01 \\
    DIN-SQL                     & 2.01 & \fst{0} & \fst{0}    & \fst{0} & \snd{0.33} & \snd{2.34} \\
    MAC-SQL                     & 4.35 & \fst{0} & \snd{0.33} & \fst{0} & 5.35       & 10.03 \\
    CHESS\textsubscript{(IR,SS,CG)} & \fst{0.67} & \fst{0} & \fst{0} & \snd{6.02} & \fst{0} & 6.69 \\
    CHESS\textsubscript{(IR,CG,UT)} & 3.01 & \fst{0} & \fst{0} & \fst{0} & 2.68       & 5.69 \\
    TA-SQL                      & 2.01 & \fst{0} & \fst{0}    & \fst{0} & \fst{0}    & \fst{2.01} \\
    GSR                         & 11.04 & \fst{0} & 0.67      & \fst{0} & \snd{0.33} & 12.04 \\
    E-SQL                       & \snd{1.34} & \fst{0} & \snd{0.33} & \fst{0} & 3.34 & 5.02 \\
    RSL-SQL                     & \fst{0.67} & \fst{0} & \fst{0}    & \fst{0} & 14.38 & 15.05 \\
    OpenSearch-SQL              & 2.01 & \fst{0} & \fst{0}    & \fst{0} & 32.44 & 34.45 \\
    \bottomrule
  \end{tabular}%
  }
\end{table}

\begin{table}
  \centering
  \caption{Error analysis on \textbf{ScienceBenchmark Dev using GPT-4o-mini} for Candidate Generation.
  The best results are in \fst{bold and underlined} while the second-best results are \snd{underlined}.
  We use \texttt{Pass@1} results for approaches producing multiple candidates.}
  \setlength\tabcolsep{2pt}
  \renewcommand{\arraystretch}{1.0}
  \label{tab:candidate_generation_error_scib_gpt4omini}
  \small
  \resizebox{0.99\columnwidth}{!}{%
  \begin{tabular}{l|c|c|c|c|c|c}
     \toprule
    \multirow{2}{*}{\makecell{Approach}}
      & \multicolumn{5}{c|}{Error Rates (\%)}
      & \multirow{2}{*}{\makecell{Total\\(\%)}} \\
    \cmidrule(lr){2-6}
      & \makecell{No Tabl/Col}
      & \makecell{No Func}
      & \makecell{Syntax Err}
      & Timeout
      & Others
      &  \\
    \midrule
    C3-SQL                      & 4.68  & \fst{0} & \fst{0} & \fst{0}    & 1.67 & 6.35 \\
    DIN-SQL                     & 8.36  & \fst{0} & \fst{0} & \snd{0.33} & \fst{0} & 8.70 \\
    MAC-SQL                     & 8.70  & \fst{0} & \fst{0} & \snd{0.33} & 5.02 & 14.05 \\
    CHESS\textsubscript{(IR,SS,CG)} & \fst{2.01} & \fst{0} & \fst{0} & \snd{0.33} & \fst{0} & \fst{2.34} \\
    CHESS\textsubscript{(IR,CG,UT)} & 12.37 & \fst{0} & \fst{0} & \fst{0}    & \fst{0} & 12.37 \\
    TA-SQL                      & 6.02  & \fst{0} & \fst{0} & \fst{0}    & \fst{0} & \snd{6.02} \\
    GSR                         & 6.35  & \fst{0} & \fst{0} & \snd{0.33} & 0.67 & 7.36 \\
    E-SQL                       & \snd{2.34} & \fst{0} & \fst{0} & \snd{0.33} & 12.71 & 15.38 \\
    RSL-SQL                     & 6.35  & \fst{0} & \fst{0} & \snd{0.33} & \fst{0} & 6.69 \\
    OpenSearch-SQL              & 10.37 & \fst{0} & \fst{0} & \snd{0.33} & \snd{0.33} & 11.04 \\
    \bottomrule
  \end{tabular}%
  }
\end{table}

\begin{figure*}[htbp]
  \centering
  \vspace{-1em}%
  \includegraphics[width=0.99\textwidth]{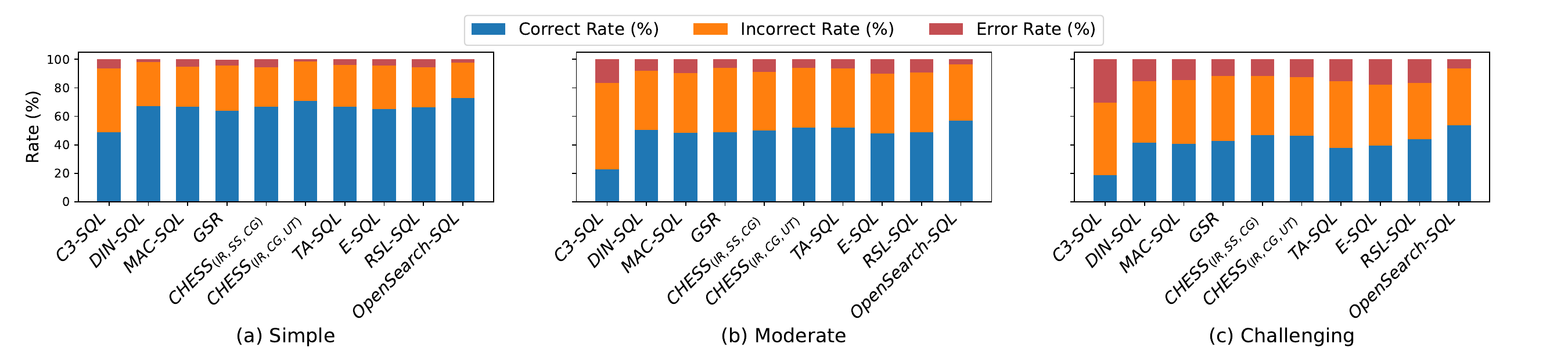}
  \vspace{-1em}%
  \caption{Correct Rates, Incorrect Rates, and Error Rates on BIRD dev set using DeepSeek-V3 for the Candidate Generation module across difficulty levels.}
  \label{fig:candidate_generation_difficulty_bird_ds}
\end{figure*}

\begin{figure*}[htbp]
  \centering
  \vspace{-1em}%
  \includegraphics[width=0.99\textwidth]{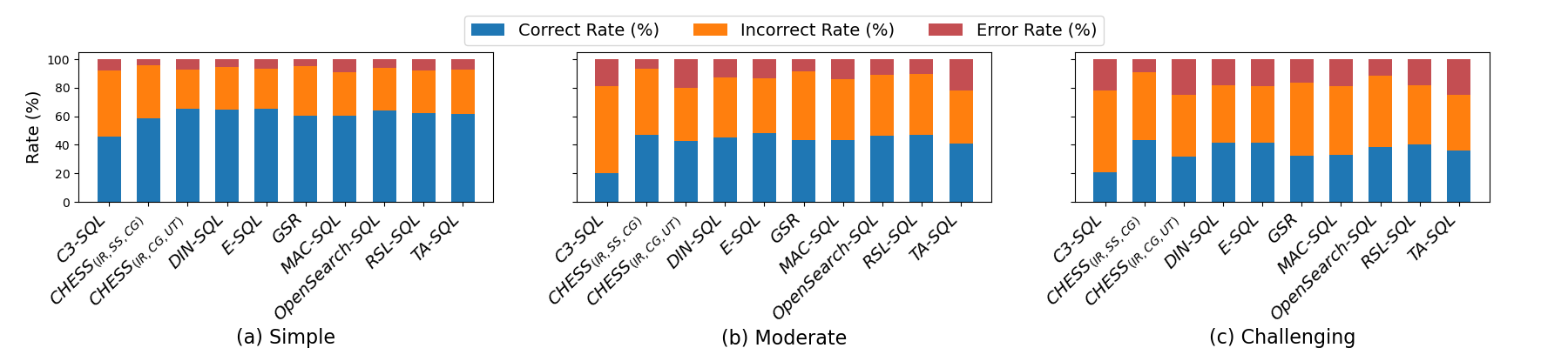}
  \vspace{-1em}%
  \caption{Correct Rates, Incorrect Rates, and Error Rates on BIRD dev set using GPT-4o-mini for the Candidate Generation module across difficulty levels.}
  \label{fig:candidate_generation_difficulty_bird_gpt}
\end{figure*}

\begin{figure*}[htbp]
  \centering
  \vspace{-1em}%
  \includegraphics[width=0.99\textwidth]{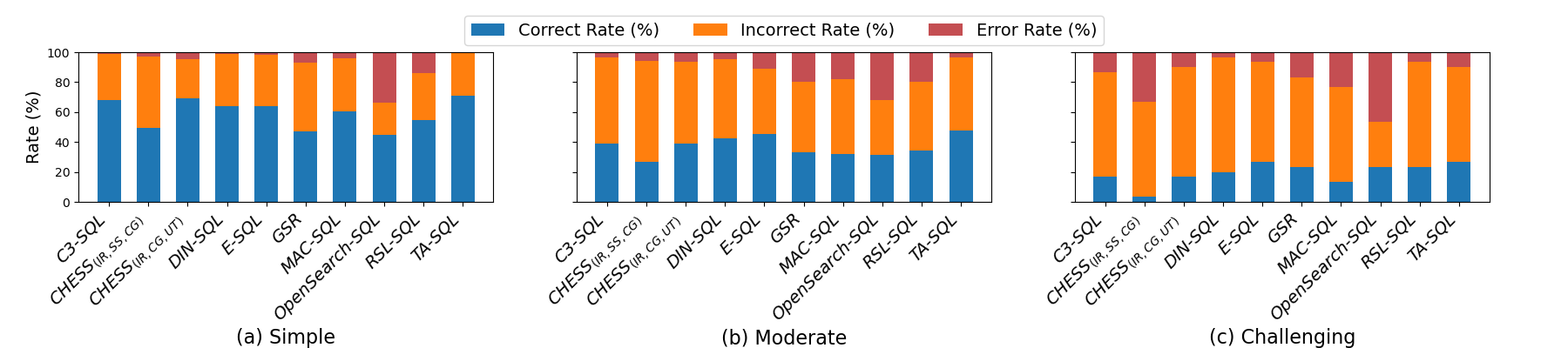}
  \vspace{-1em}%
  \caption{Correct Rates, Incorrect Rates, and Error Rates on ScienceBenchmark dev set using DeepSeek-V3 for the Candidate Generation module across difficulty levels.}
  \label{fig:candidate_generation_difficulty_sb_ds}
\end{figure*}

\begin{figure*}[htbp]
  \centering
  \vspace{-1em}%
  \includegraphics[width=0.99\textwidth]{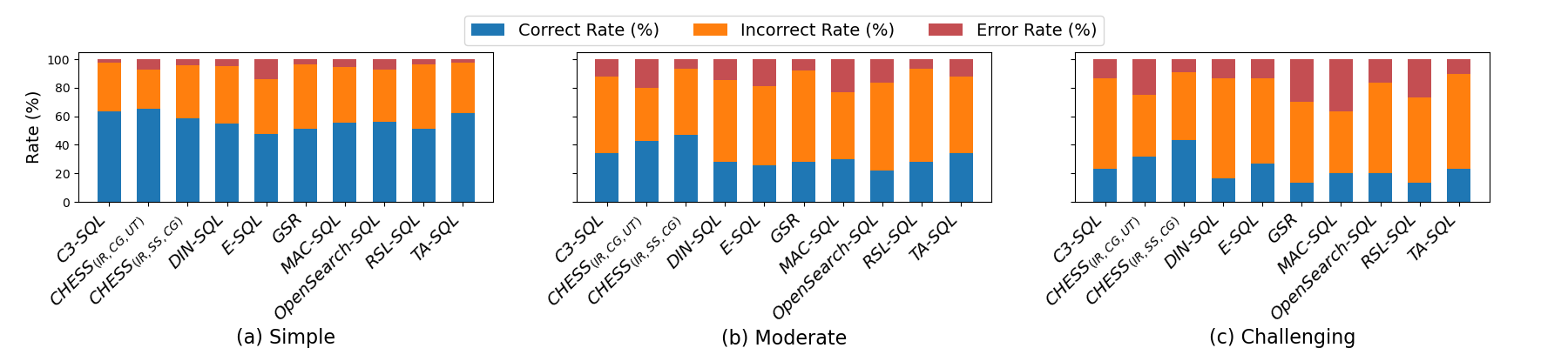}
  \vspace{-1em}%
  \caption{Correct Rates, Incorrect Rates, and Error Rates on ScienceBenchmark dev set using GPT-4o-mini for the Candidate Generation module across difficulty levels.}
  \label{fig:candidate_generation_difficulty_sb_gpt}
\end{figure*}

\subsubsection{Result analysis on difficulty level.}
\label{appendix:candidate_generation_difficulty}

Figure \ref{fig:candidate_generation_difficulty_bird_ds} \ref{fig:candidate_generation_difficulty_bird_gpt} \ref{fig:candidate_generation_difficulty_sb_ds} \ref{fig:candidate_generation_difficulty_sb_gpt} present the \emph{Correct Rates}, \emph{Incorrect Rates}, and \emph{Error Rates} for the Candidate Generation module across three difficulty levels—Simple, Moderate, and Challenging—for each evaluated approach under the four experiment settings.

\textbf{Overall trend.} As we can see, when the difficulty level increases gradually, the Correct Rates show a declining trend, while IR increases and becomes the dominant mass of errors; ER remains comparatively small but shows a mild uptick on Challenging. This confirms that most failures at the greater difficulty are semantically wrong yet executable SQL (e.g., wrong join path, boundary/filter mismatch, misused grouping), rather than parser/dialect errors.

\textbf{Dataset effect.} Within each difficulty band, ScienceBenchmark yields lower CR and higher IR than BIRD for many methods. The gap widens on Challenging, reflecting ScienceBenchmark’s denser schemas.
Some systems (e.g., OpenSearch-SQL) have done targeted alignment for the BIRD dataset, thus show good results on BIRD, while on ScienceBenchmark, their advantage narrows. Conversely, concise pipelines remain comparatively stable across backbones.

\textbf{Backbone LLM effect.} Changing backbone LLMs reorders methods within each difficulty band, but it does not change the global trend: CR decreases, and IR increases with difficulty in both LLMs.

\subsection{Additional Results for Query Revision Module}

\subsubsection{Result analysis on difficulty level.}
\label{appendix:revision_difficulty}
\begin{figure*}[htbp]
  \centering
  \vspace{-1em}%
  \includegraphics[width=0.99\textwidth]{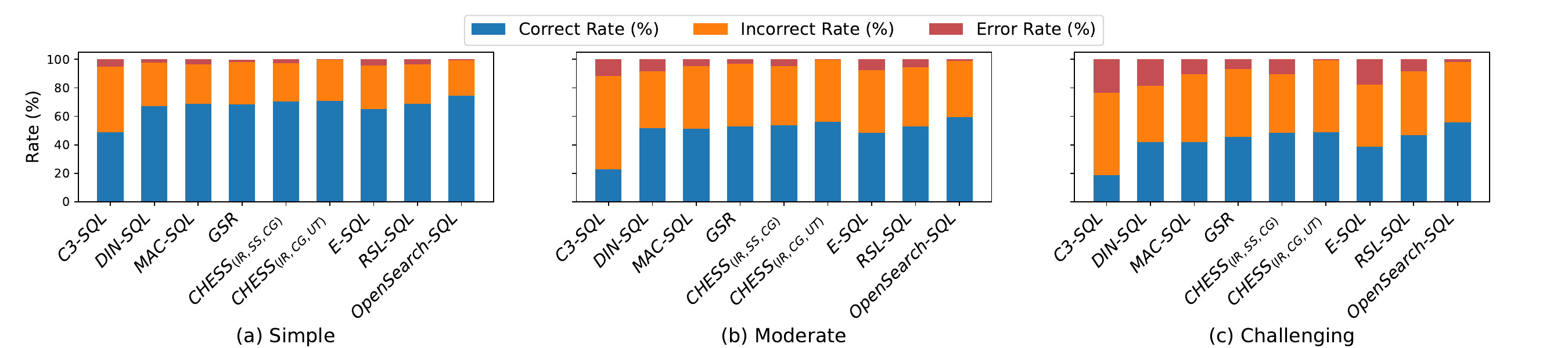}
  \vspace{-1em}%
  \caption{Correct Rates, Incorrect Rates, and Error Rates on BIRD dev set using DeepSeek-V3 for Query Revision module across difficulty levels.}
  \label{fig:correct_rate_analysis_revision_bird_ds}
\end{figure*}

\begin{figure*}[htbp]
  \centering
  \vspace{-1em}%
  \includegraphics[width=0.99\textwidth]{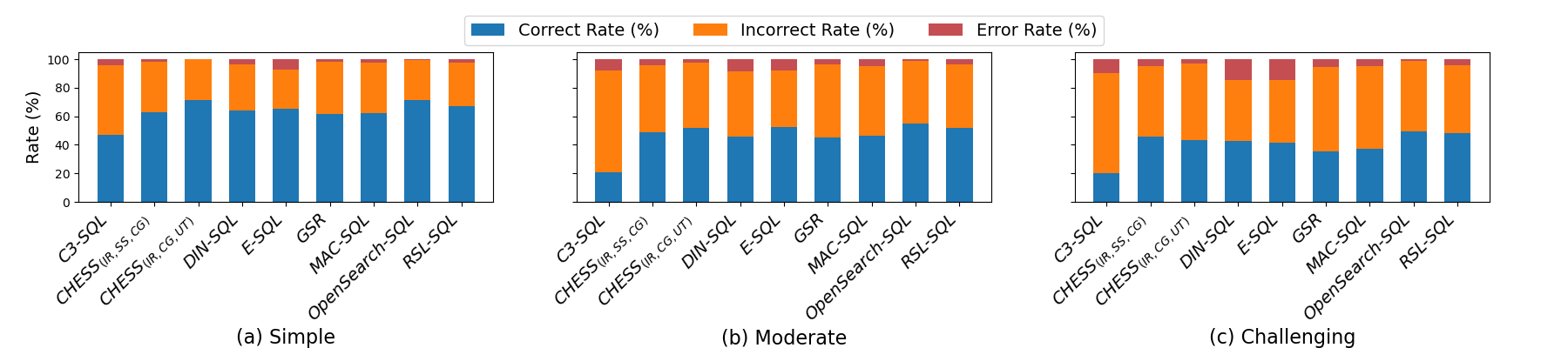}
  \vspace{-1em}%
  \caption{Correct Rates, Incorrect Rates, and Error Rates on BIRD dev set using GPT-4o-mini for Query Revision module across difficulty levels.}
  \label{fig:correct_rate_analysis_revision_bird_gpt}
\end{figure*}

\begin{figure*}[htbp]
  \centering
  \vspace{-1em}%
  \includegraphics[width=0.99\textwidth]{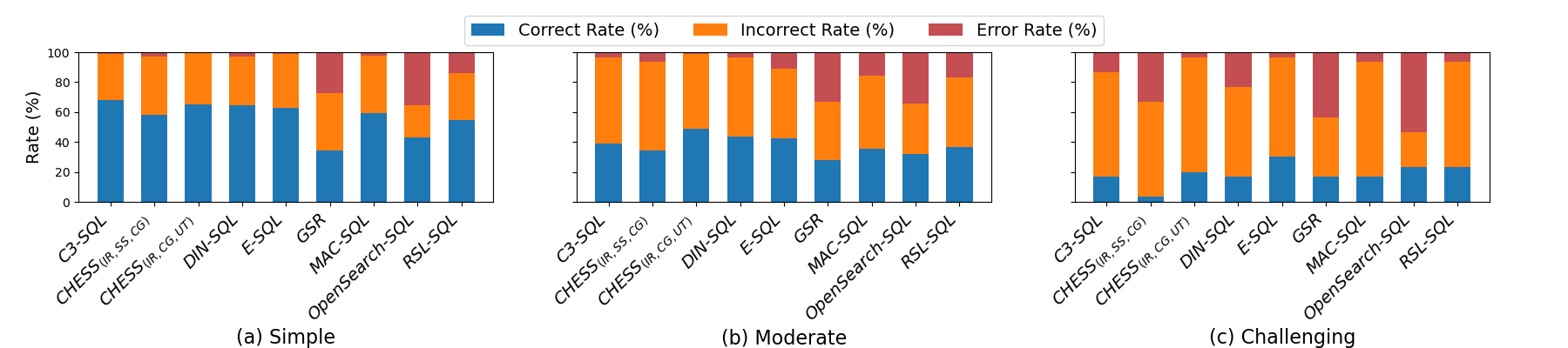}
  \vspace{-1em}%
  \caption{Correct Rates, Incorrect Rates, and Error Rates on ScienceBenchmark dev set using DeepSeek-V3 for Query Revision module across difficulty levels.}
  \label{fig:correct_rate_analysis_revision_sb_ds}
\end{figure*}

\begin{figure*}[htbp]
  \centering
  \vspace{-1em}%
  \includegraphics[width=0.99\textwidth]{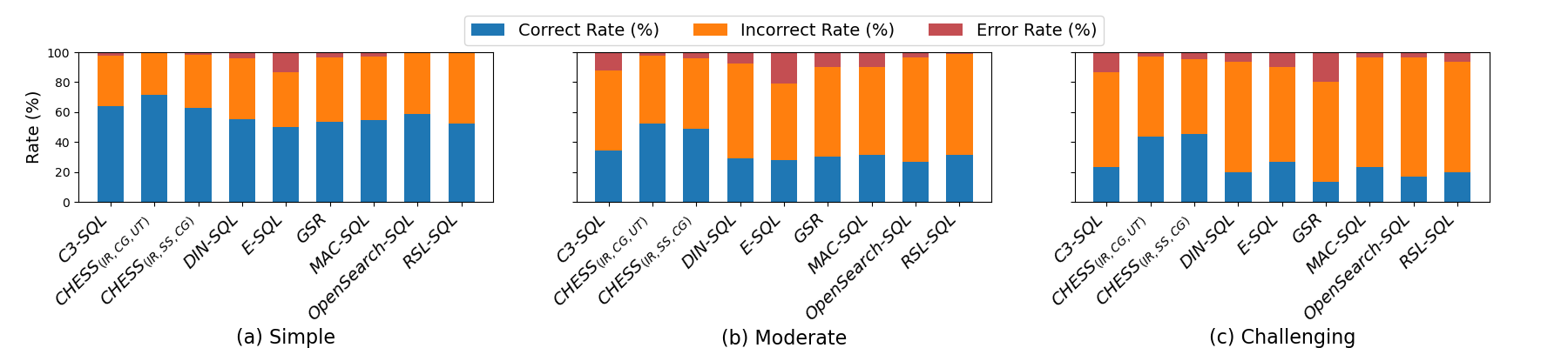}
  \vspace{-1em}%
  \caption{Correct Rates, Incorrect Rates, and Error Rates on ScienceBenchmark dev set using GPT-4o-mini for Query Revision module across difficulty levels.}
  \label{fig:correct_rate_analysis_revision_sb_gpt}
\end{figure*}

Figure \ref{fig:correct_rate_analysis_revision_bird_ds}, Figure \ref{fig:correct_rate_analysis_revision_bird_gpt}, Figure \ref{fig:correct_rate_analysis_revision_sb_ds} and Figure \ref{fig:correct_rate_analysis_revision_sb_gpt} present the \emph{Correct Rates}, \emph{Incorrect Rates}, and \emph{Error Rates} across three difficulty levels for each approach after applying the Query Revision module.

\textbf{Overall trend.}
Similar to the trend previously, the correct rate gradually decreases as query difficulty progresses from Simple to Challenging.
Notably, Incorrect queries continue to constitute the majority of unsuccessful results, maintaining relatively high Incorrect Rates post-revision. These queries are syntactically valid yet fail to reproduce results identical to the ground truth SQL, revealing a persistent semantic mismatch. Their prevalence, therefore, indicates that current revision strategies are still insufficient for faithfully capturing and accurately aligning with the semantic intentions expressed in users' NL queries.

\textbf{Dataset effect.} On BIRD, several systems preserve a reasonable CR margin from Simple to Moderate and achieve visible IR reduction after revision. On ScienceBenchmark, however, the CR drop is not as obvious as BIRD. Notably, OpenSearch-SQL, which has undergone extensive alignment optimization for BIRD, does not transfer well to ScienceBenchmark: its CR advantage narrows, indicating that dataset-specific alignment does not generalize and must be re-designed for new schema distributions.

\textbf{Backbone LLM effect.} Changing the backbone reorders methods within each difficulty band but does not alter the global shape.

\textit{\textbf{Insight:} Current query revision approaches remain inadequate at identifying and rectifying Incorrect queries, which represent the predominant error type.
Since these queries are syntactically correct, they are more subtle and challenging to detect, reflecting a persistent misalignment with users' intended semantics.
}

\textbf{Practical guide.} Boosting CR requires semantic alignment rather than additional surface rewrites. Alignment strategies must generalize across datasets to ensure robustness and generalization.

\begin{figure}[htbp]
  \centering
    \begin{subfigure}[b]{0.99\linewidth}
    \centering
    \includegraphics[width=\linewidth]{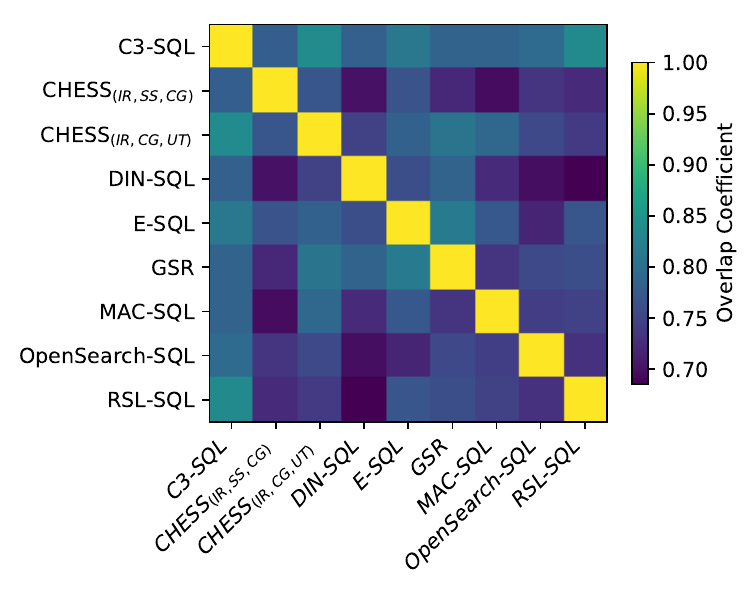}
    \caption{Overlap on the BIRD (using GPT-4o-mini)}
    \label{fig:overlap_a}
  \end{subfigure}
  \hfill
  \begin{subfigure}[b]{0.99\linewidth}
    \centering
    \includegraphics[width=\linewidth]{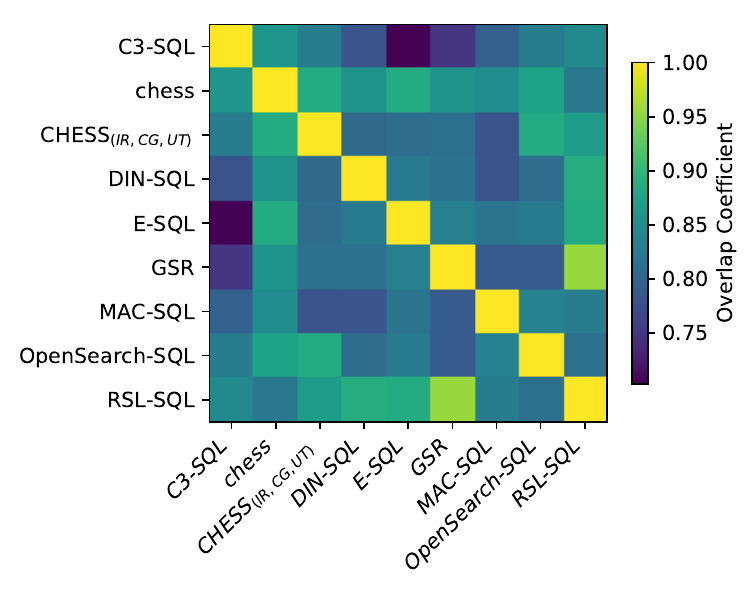}
    \caption{Overlap on the ScienceBenchmark (using GPT-4o-mini)}
    \label{fig:overlap_b}
  \end{subfigure}
  \caption{The coefficient heatmap of different solutions on Incorrect cases (using GPT-4o-mini).}
  \label{fig:incorrect_overlap_gpt_4o}
\end{figure}

\subsubsection{An in-depth analysis of incorrect cases}
\label{appendix:revision_in_depth_analysis}
Similar to the results shown in Section \ref{sec:incorrect_overlap}, we observed an unexpectedly high degree of overlap in incorrect queries generated by different methods using GPT-4o-mini on the two datasets, as illustrated in Figure \ref{fig:incorrect_overlap_gpt_4o}.

\subsubsection{Detailed analysis on Query Revision performance.}
\label{appendix:revision_in_depth_analysis_full}
We present the full results of Query Revision module using Correct-Improvement (CI), Incorrect-to-Correct (I2C), and Error-to-Correct (E2C), Correct-to-Incorrect(C2I), Correct-to-Error(C2E) along with efficiency (\#tokens/\#LLM calls), as shown in Table \ref{tab:revision_effectiveness_bird_deepseek}, Table \ref{tab:revision_effectiveness_bird_gpt4omini}, Table \ref{tab:revision_effectiveness_scibench_deepseek} and Table \ref{tab:revision_effectiveness_scibench_gpt4omini}.

\begin{table}[h]
\small
\setlength\tabcolsep{1pt}
\renewcommand\arraystretch{1.0}
\caption{Effectiveness and efficiency on Query Revision on BIRD dev with DeepSeek-V3. The best results are in \fst{bold and underlined} while the second-best results are \snd{underlined}.}
\vspace{-0.5em}
\setlength\tabcolsep{2pt}
\resizebox{0.99\columnwidth}{!}{%
\centering \begin{tabular}{l|r|r|r|r|r|r|r}
\toprule Approach & \makecell{CI\\(\%)} & \makecell{I2C\\(\%)} & \makecell{E2C\\(\%)} & \makecell{C2I\\(\%)} & \makecell{C2E\\(\%)} & \makecell{\#Tok-\\ens} & \makecell{\#LLM\\Calls} \\ \midrule
C3-SQL & -0.51 & 1.43 & 4.97 & 3.93 & \fst{0.00} & -- & -- \\
DIN-SQL & 0.66 & 2.02 & 21.33 & 0.98 & 1.31 & 4948 & \fst{1} \\
MAC-SQL & 3.55 & 3.66 & 13.16 & \snd{0.22} & \fst{0.00} & \fst{1566} & \snd{2} \\
CHESS\textsubscript{(IR,SS,CG)} & \snd{6.11} & 6.89 & 29.36 & 0.65 & \snd{0.55} & \snd{1844} & \fst{1} \\
CHESS\textsubscript{(IR,CG,UT)} & 2.18 & \fst{15.16} & \fst{47.54} & 8.81 & \fst{0.00} & 106126 & 21 \\
GSR & \fst{7.27} & \snd{9.49} & \snd{38.10} & 2.16 & 0.34 & 2220 & \snd{2} \\
E-SQL & 0.00 & 3.53 & 14.04 & 3.06 & 0.91 & 24924 & \snd{2} \\
RSL-SQL & 4.97 & 4.69 & 17.95 & \fst{0.00} & \fst{0.00} & 3315 & \snd{2} \\
OpenSearch-SQL& 3.16 & 8.67 & 34.04 & 2.47 & \fst{0.00} & 6652 & 4.6\\ \bottomrule \end{tabular} } \label{tab:revision_effectiveness_bird_deepseek} \end{table}

\begin{table}[h]
\small
\setlength\tabcolsep{1pt}
\renewcommand\arraystretch{1.0}
\caption{Effectiveness and efficiency on Query Revision on BIRD dev with GPT-4o-mini. The best results are in \fst{bold and underlined} while the second-best results are \snd{underlined}.}
\vspace{-0.5em}
\setlength\tabcolsep{2pt}
\resizebox{0.99\columnwidth}{!}{%
\centering
\begin{tabular}{l|r|r|r|r|r|r|r}
\toprule
Approach & \makecell{CI\\(\%)} & \makecell{I2C\\(\%)} & \makecell{E2C\\(\%)} & \makecell{C2I\\(\%)} & \makecell{C2E\\(\%)} & \makecell{\#Tok-\\ens} & \makecell{\#LLM\\Calls} \\ \midrule
C3-SQL                  &  2.56                   &  2.52                   &  8.33                   &  4.02 &  \fst{0}    & --            & -- \\
DIN-SQL                 & -0.11                   &  4.16                   & 19.12                   &  4.37 &  1.27 & 4791.23                  & \fst{1} \\
MAC-SQL                 &  4.45                   &  1.45                   & 21.84                   &  \fst{1.24} &  \fst{0}    & \snd{1729.12}                  & \snd{2} \\
CHESS\textsubscript{(IR,SS,CG)} &  6.72         &  10.49                  & 35.29                   &  4.64 &  0.37 & \fst{1683.28}            & \fst{1} \\
CHESS\textsubscript{(IR,CG,UT)} & \snd{14.04}    & \snd{16.09}             & \snd{38.38}             &  4.26 &  \snd{0.12} & 115055.19                & 23.19 \\
GSR                      &  2.6                 &  4.85                   & 21.3                   &  3.71 &  0.25 & 2448.84                  & \snd{2} \\
E-SQL                    &  1.91                  &  7.52                   & 28.29                   &  4.04 &  3.03 & 25022.94                 & \snd{2} \\
RSL-SQL                  &  9.15                  & 10.06                   & 35.86                   &  \snd{2.82} &  0.47 & 2802.02                  & \snd{2} \\
OpenSearch-SQL           & \fst{14.07}            & \fst{17.71}             & \fst{51.20}             &  4.38 &  \fst{0}    & 12968.94                 & 6.64 \\
\bottomrule
\end{tabular}
}
\label{tab:revision_effectiveness_bird_gpt4omini}
\end{table}

\begin{table}[h]
\small
\setlength\tabcolsep{1pt}
\renewcommand\arraystretch{1.0}
\caption{Effectiveness and efficiency on Query Revision on ScienceBenchmark dev with DeepSeek-V3. The best results are in \fst{bold and underlined} while the second-best results are \snd{underlined}.}
\vspace{-0.5em}
\setlength\tabcolsep{2pt}
\resizebox{0.99\columnwidth}{!}{%
\centering
\begin{tabular}{l|r|r|r|r|r|r|r}
\toprule
Approach & \makecell{CI\\(\%)} & \makecell{I2C\\(\%)} & \makecell{E2C\\(\%)} & \makecell{C2I\\(\%)} & \makecell{C2E\\(\%)} & \makecell{\#Tok-\\ens} & \makecell{\#LLM\\Calls} \\ \midrule
C3-SQL                  & 0                       & 0                       & 0                       & \fst{0}    & \fst{0}    & --                  & -- \\
DIN-SQL                 & 1.27                    & 2.99                    & 14.29                   & \snd{0.63} & 1.27 & 4349.08         & \fst{1} \\
MAC-SQL                 & 1.42                   & 2.34                    & 10                 & 2.84 & \fst{0}    & 2395.41                   & \snd{2} \\
CHESS\textsubscript{(IR,SS,CG)} & \fst{19.47}     & \fst{14.55}             & 0                   & 1.77 & \fst{0}    & \snd{2319.17}             & \fst{1} \\
CHESS\textsubscript{(IR,CG,UT)} & \snd{1.83}      & 9.32              & \fst{52.94}             & 10.37& \fst{0}    & 201372.44                 & 23.42 \\
GSR                      & -24.8                 & \snd{14.08}      & \snd{33.33}                   & 23.14 & 28.1 & \fst{1756.85}             & \snd{2} \\
E-SQL                    & -2.45                  & 4.96                    & 13.33                   & 6.13 & \snd{1.23} & 15088.02                  & \snd{2} \\
RSL-SQL                  & 1.47                   & 0.00                    & 4.44                    & \fst{0} & \fst{0}    & 6616.00                   & \snd{2} \\
OpenSearch-SQL           & -1.74                  & 7.41                    & 3.88                 & 4.35 & 6.09 & 18574.91                  & 7.17 \\
\bottomrule
\end{tabular}
}
\label{tab:revision_effectiveness_scibench_deepseek}
\end{table}

\begin{table}[h]
\small
\setlength\tabcolsep{1pt}
\renewcommand\arraystretch{1.0}
\caption{Effectiveness and efficiency on Query Revision on ScienceBenchmark dev with GPT-4o-mini. The best results are in \fst{bold and underlined} while the second-best results are \snd{underlined}.}
\vspace{-0.5em}
\setlength\tabcolsep{2pt}
\resizebox{0.99\columnwidth}{!}{%
\centering
\begin{tabular}{l|r|r|r|r|r|r|r}
\toprule
Approach & \makecell{CI\\(\%)} & \makecell{I2C\\(\%)} & \makecell{E2C\\(\%)} & \makecell{C2I\\(\%)} & \makecell{C2E\\(\%)} & \makecell{\#Tok-\\ens} & \makecell{\#LLM\\Calls} \\ \midrule
C3-SQL                  & 0                       & 0                       & 0                       & \fst{0}    & \fst{0}    & --                  & -- \\
DIN-SQL                 & 2.34                    & 1.38                    & 15.38                   & \snd{1.56} & \snd{0.78} & 4156.48               & \fst{1} \\
MAC-SQL                 & 0.76                    & 3.20                    &  9.52                   & 5.30 & \fst{0}    & 2675.69                   & \snd{2} \\
CHESS\textsubscript{(IR,SS,CG)} & \fst{26.32}     & \fst{13.30}             & 14.29             & 2.11 & \fst{0}    & \snd{1683.28}             & \fst{1} \\
CHESS\textsubscript{(IR,CG,UT)} & \snd{20.49}     & \snd{12.14}             & \fst{43.24}             & 6.56 & \fst{0}    & 115055.19                 & 23.19 \\
GSR                      & 4.96                   & 5.13                    & 18.18                   & 4.13 & 0.83 & \fst{1498.20}                   & \snd{2} \\
E-SQL                    & 5.17                   & 6.57                    & 4.35          & 2.59 & 1.72 & 21360.30                  & \snd{2} \\
RSL-SQL                  & 5.79                   & 0.00                    & 35         & \fst{0} & \fst{0} & 3720.00                & \snd{2} \\
OpenSearch-SQL           & 6.35                   & 7.14                    & \snd{33.33}                   & 10.32& \fst{0}    & 34752.35                  & 11.33 \\
\bottomrule
\end{tabular}
}
\label{tab:revision_effectiveness_scibench_gpt4omini}
\end{table}

\subsection{Example of New Discoveries on Inaccurate Gold SQL Queries and Evaluations}
\label{appendix:incorrect_gold_sql_example}

During our evaluation of the BIRD dataset, we identified several gold SQL queries that are inaccurate and have not been discovered previously by existing works, such as \cite{liu2025nl2sql-bugs}. One example of our findings is the question: \emph{"Name the foreign name of the card that has an abzan watermark? List out the type of this card"} (BIRD question\_id = 448) is currently aligned with an \emph{incorrect} gold SQL query:

\begin{adjustwidth}{1em}{0em}
\texttt{SELECT DISTINCT {\color{red} \underline{T1.name}}, T1.type FROM cards AS T1 INNER JOIN foreign\_data AS T2 ON T2.uuid = T1.uuid WHERE T1.watermark = 'abzan';}
\end{adjustwidth}


In the above query, the column \texttt{T1.name} is incorrectly selected. The correct SQL query should instead select \texttt{T2.name} like below:

\begin{adjustwidth}{1em}{0em}
\texttt{SELECT DISTINCT {\color{blue} \underline{T2.name}}, T1.type FROM cards AS T1 INNER JOIN foreign\_data AS T2 ON T2.uuid = T1.uuid WHERE T1.watermark = 'abzan';}
\end{adjustwidth}

Another example of a provided inaccurate query is shown below.

\textit{Question:}
Name all cards with a 2015 frame style ranking below 100 on EDHRec?

\begin{itemize}[label=$\vartriangleright$, nosep, noitemsep, left=0.5em]
    \item \textit{Incorrect Gold SQL query:}\hspace{1.1em} {\small{\texttt{SELECT {\color{red} \underline{id}} FROM cards WHERE edhrecRank < 100 AND frameVersion = 2015;}}}

    \item \textit{Correct SQL query:}\hspace{3.9em} {\small{\texttt{SELECT {\color{blue} \underline{name}} FROM cards WHERE edhrecRank < 100 AND frameVersion = 2015;}}}
\end{itemize}

\textbf{Over-Strict Evaluation Rule}. Consider the question: \emph{"Which user added a bounty amount of 50 to the post title mentioning variance?"} (BIRD question\_id = 586).
BIRD's Gold SQL query returns both the \texttt{DisplayName} and \texttt{Title} columns. An alternative query that retrieves only the \texttt{DisplayName} column still satisfies most information needs. However, BIRD's existing evaluation rules label it as incorrect.


The correct SQL query should return the \texttt{name} column, but the given gold SQL only returns the \texttt{id} column.
Given that the gold SQL serves as the critical reference for evaluating the correctness and effectiveness of NL2SQL methods, inaccuracies in these gold queries can significantly undermine the validity and reliability of the benchmark results.

\textbf{Semantic Ambiguity.} Consider the question: \emph{"Which school in Contra Costa has the highest number of test takers?"} (BIRD question\_id = 22).

In the underlying schema, a \emph{“school”} can be represented by any of the columns \texttt{CDSCode}, \texttt{sname}, \texttt{School Name}, or \texttt{School}. Each of these columns can be reasonably regarded as a valid answer, highlighting the inherent ambiguity in the question formulation. Because each of these columns is a plausible surrogate for the same concept, multiple SQL queries—returning different but semantically equivalent result sets—should be deemed correct. The current one-to-one evaluation paradigm of BIRD cannot capture this variability and thus may unfairly penalize systems that choose an alternative legitimate representation.

\textbf{Practical Guide.} we provide the following recommendations for practitioners and benchmark developers.

First, to mitigate the impact of inaccurate gold annotations, we recommend implementing multi-reference evaluation where generated queries are validated against multiple acceptable SQL variants rather than a single gold query—this can be achieved through result-equivalence checking (comparing execution results across multiple semantically equivalent queries) or through LLM-based semantic equivalence verification. For production systems, establish a human-in-the-loop validation process for queries that execute successfully but are marked incorrect by automated evaluation, as these cases often indicate annotation errors rather than system failures.

Second, to address strict evaluation rules that fail to recognize semantically equivalent results, adopt flexible evaluation metrics such as partial credit scoring (rewarding queries that return a superset or subset of required columns) or schema-normalized comparison (treating different valid representations of the same entity as equivalent). When deploying NL2SQL systems, provide users with query explanation mechanisms that clarify what information the generated SQL retrieves, enabling users to verify semantic correctness even when result formatting differs from expectations.

Third, to handle semantic ambiguity in natural language questions, implement ambiguity detection and clarification mechanisms: use LLM-based analysis to identify potentially ambiguous terms (e.g., "school" could map to multiple columns), generate clarification questions for users, or provide multiple query interpretations with explanations. For benchmark developers, we strongly advocate for systematic quality assurance protocols including: (1) multi-annotator consensus requirements for gold SQL creation, (2) automated consistency checking to detect annotation errors (e.g., mismatches between questions asking for "names" and gold queries returning IDs), and (3) explicit ambiguity annotations marking questions with multiple valid interpretations.

Furthermore, we recommend developing semantically-aware evaluation frameworks that can recognize and appropriately credit semantically equivalent but syntactically different SQL queries, moving beyond simple string matching or single-reference execution accuracy. These quality improvements are essential not only for fair benchmark evaluation but also for training more robust NL2SQL systems, as models trained or evaluated on noisy benchmarks may learn to replicate annotation errors rather than develop genuine semantic understanding.

\subsection{Case Study on Leveraging {\toolname} for NL2SQL System Development}
\label{appendix:case_study}
This section serves as the case study of the usability and recommendations on leveraging {\toolname} for NL2SQL system development. We follow the systematic workflow proposed in Sec \ref{usability}.

\textit{First, establish a performance profile by running {\toolname} with different baselines on your target dataset to generate a comprehensive performance profile across all three core modules. This profile identifies the primary bottleneck.}
For Example, if the Schema Selection F1-score < 75\%, schema linking represents the critical constraint; if the Candidate Generation Error Rate > 10\%, focus on reducing execution errors through better prompting or intermediate representations; if the Query Revision Correctness Improvement < 5\% despite high Error-to-Correct rates, the revision module may be ineffective at addressing the dominant Incorrect query problem.

\textit{Second, apply targeted optimizations to the bottleneck module while monitoring cross-module impacts.}
For instance, improving schema precision from 32\% to 85\% not only enhances Schema Selection F1-score but also simplifies downstream candidate generation by reducing irrelevant context, potentially enabling faster generation strategies. Use {\toolname} to measure these cascading effects and validate that optimizations to one module do not inadvertently degrade others.

\textit{Third, conduct cost-benefit analysis at the module level using our effectiveness and efficiency metrics.}
Compare approaches like CHESS\textsubscript{(IR,SS,CG)} (63.43\% accuracy, 310K tokens, 80 LLM calls) against MAC-SQL (60.82\% accuracy, 7.5K tokens, 4 LLM calls) to determine whether marginal accuracy gains justify 40× cost increases for your specific application requirements. Define clear operating constraints—such as maximum latency budget (e.g., < 500ms end-to-end), cost budget (e.g., < \$0.01 per query), or minimum acceptable accuracy (e.g., > 60\% execution accuracy)—and use {\toolname}'s fine-grained metrics to navigate the accuracy-efficiency trade-off space systematically.
For example, latency-critical applications should favor approaches with minimal LLM calls ($\leq$ 5), while accuracy-critical applications may allocate larger computational budgets to high-precision schema selection and multi-candidate generation

\textit{Finally, leverage {\toolname} for A/B testing and continuous improvement.}
When evaluating alternative strategies for a specific module (e.g., Few-Shot CoT vs. Preliminary-SQL for schema selection), use the framework's standardized metrics to make evidence-based decisions, and continuously benchmark production systems to detect performance degradation over time.

By treating NL2SQL development as a modular optimization problem where each component can be independently analyzed, improved, and composed, practitioners can systematically construct systems that optimally balance accuracy, efficiency, and deployment constraints, thus transforming {\toolname} from a benchmarking tool into a practical framework for iterative system development.

\subsection{Additional Related Work}
\label{appendix:related_work}
DIN-SQL~\cite{pourreza2024din} splits the task into schema linking, classification, SQL generation, and self-correction, enabling LLMs to produce more precise SQL queries. MAC-SQL~\cite{macsql2025} decomposes the question into a series of sub-questions and generates corresponding sub-queries before generating the final SQL. CHESS~\cite{talaei2024chess} proposes harnessing contextual information by efficient retrieval and schema pruning methods to enhance SQL generation. TA-SQL~\cite{qu2024before} introduces alignment-based strategies to mitigate hallucinations in SQL generation. CHASE-SQL~\cite{pourreza2025chasesql} enhances the reasoning process of SQL generation by multiple CoT strategies. OpenSearch-SQL~\cite{xie2025opensearchsql} uses dynamic few-shot and multiple alignment mechanisms to enhance the performance.
ROUTE~\cite{qin2025route} proposes a multitask tuning paradigm and multitask collaboration strategy for NL2SQL.

\subsection{Interactive Perspective Analysis}
\label{appendix:interactive_analysis}
We provide some case studies of the analysis of error propagation and interaction relationships between the Schema Selection module and Candidate Generation module using {\toolname}.

\subsubsection{Case Study 1:}

Table \ref{tab:correct_rate_bird_deepseekv3_by_recall} breaks downstream correctness by whether the schema stage attains full recall (all required tables/columns selected; “Recall = 1”) or misses at least one (“Recall < 1”). Two clear patterns emerge.

\textit{The absence of the requisite schema results in a substantial decrease in performance within downstream modules.}
Across all methods, correct rates collapse when recall falls below 1. The drops are large—typically 15–30 percentage points—and hold for both CG and QR. For example, DIN-SQL falls from 69.22\% (Recall=1) to 39.10\% (Recall<1) at CG, and from 69.32\% to 40.12\% at QR; MAC-SQL drops from 60.66\% to 25.88\% at CG (QR: 62.46\%→32.94\%); OpenSearch-SQL (Pass@20) moves from 73.22\% to 54.92\% (QR: 69.48\%→53.28\%). These gaps show error propagation: once the schema stage fails to surface a needed table/column (recall deficit), CG composes SQL queries with incomplete parts, and QR seldom recovers—its edits cannot introduce missing entities that were never selected.

\textit{More candidates help, but cannot repair missing schema.}
Within a fixed recall stratum, Pass@k monotonically raises CG correctness (e.g., OpenSearch-SQL grows from 67.63\% at k=1 to 73.22\% at k=20 when Recall=1, and 48.36\%→54.92\% when Recall<1). Yet the Recall=1 vs. Recall<1 gap persists even at k=20, indicating that selection/repair cannot overcome schema omissions.

\begin{table}
  \centering
  \caption{Correct-rate analysis on \textbf{BIRD Dev + DeepSeek-V3} conditioned on schema recall.
  We report percentages (higher is better).}
  \label{tab:correct_rate_bird_deepseekv3_by_recall}
  \setlength\tabcolsep{2pt}
  \renewcommand{\arraystretch}{1.0}
  \small
  \resizebox{\columnwidth}{!}{%
  \begin{tabular}{l|cc|cc}
    \toprule
    \multirow{2}{*}{\makecell{Approach}} &
      \multicolumn{2}{c|}{\makecell{Candidate Generation\\Correct Rate (\%)}} &
      \multicolumn{2}{c}{\makecell{Query Revision\\Correct Rate (\%)}} \\
    \cmidrule(lr){2-3}\cmidrule(lr){4-5}
      & \makecell{Recall = 1}
      & \makecell{Recall $<1$}
      & \makecell{Recall = 1}
      & \makecell{Recall $<1$} \\
    \midrule
    C3\mbox{-}SQL (Pass@1)          & 42.51\% & 15.10\% & 42.28\% & 15.10\% \\
    C3\mbox{-}SQL (Pass@5)          & 45.31\% & 16.33\% & 42.28\% & 15.10\% \\
    C3\mbox{-}SQL (Pass@10)         & 46.16\% & 16.33\% & 42.28\% & 15.10\% \\
    C3\mbox{-}SQL (Pass@15)         & 46.47\% & 17.55\% & 42.28\% & 15.10\% \\
    C3\mbox{-}SQL (Pass@20)         & 46.86\% & 18.78\% & 42.28\% & 15.10\% \\
    DIN\mbox{-}SQL                  & 69.22\% & 39.10\% & 69.32\% & 40.12\% \\
    MAC\mbox{-}SQL                  & 60.66\% & 25.88\% & 62.46\% & 32.94\% \\
    CHESS (IR, SS, CG)              & 68.51\% & 44.42\% & 71.27\% & 49.64\% \\
    TA\mbox{-}SQL                   & 66.06\% & 40.98\% & --       & --      \\
    GSR                             & 66.67\% & 40.52\% & 70.30\% & 45.67\% \\
    RSL\mbox{-}SQL                  & 62.57\% & 45.05\% & 65.85\% & 46.65\% \\
    OpenSearch\mbox{-}SQL (Pass@1)  & 67.63\% & 48.36\% & 69.48\% & 53.28\% \\
    OpenSearch\mbox{-}SQL (Pass@5)  & 71.03\% & 52.46\% & 69.48\% & 53.28\% \\
    OpenSearch\mbox{-}SQL (Pass@10) & 72.17\% & 53.80\% & 69.48\% & 53.28\% \\
    OpenSearch\mbox{-}SQL (Pass@15) & 72.66\% & 54.10\% & 69.48\% & 53.28\% \\
    OpenSearch\mbox{-}SQL (Pass@20) & 73.22\% & 54.92\% & 69.48\% & 53.28\% \\
    \bottomrule
  \end{tabular}%
  }
\end{table}

\subsubsection{Case Study 2:}
Table \ref{tab:recall_bird_deepseekv3} reports schema selection recall under four outcome conditions: when the Candidate Generation (CG) module produces a correct vs. wrong SQL, and when the Query Revision (QR) module produces a correct vs. wrong SQL. We also list multi-candidate methods at different \texttt{Pass@k} values.

\textit{Higher schema recall strongly correlates with success downstream.}
For nearly all approaches, the recall measured on queries that end up correct is consistently higher than the recall measured on queries that end up wrong.
This pattern indicates error propagation: if the schema selection stage omits a needed table or column, CG is forced to compose with incomplete parts, and QR rarely recovers because the missing elements were never surfaced. In short, low early recall makes later stages unrecoverable.

\textit{Two distinct failure regimes appear.}
A contrasting behavior is visible for breadth-heavy systems (e.g., \textsc{OpenSearch-SQL}): even when the final SQL is wrong, schema recall remains very high (0.95), only a few points below the “correct” condition (0.98). Here, the schema stage over-selects (high recall, lower precision), so errors arise downstream from semantic composition—wrong join path, mis-scoped filters, or aggregations—not from missing schema items. By comparison, structured pipelines such as \textsc{CHESS}\textsubscript{(IR,SS,CG)} show larger recall drops on wrong cases (e.g., 0.92→0.82), which is a recall-limited regime: missing tables/columns at the schema stage directly cause downstream failure.
\textbf{Implication:} mitigation must match regime—improve recall (question-aware retrieval, bridge-table priors) for recall-limited systems; improve precision and reasoning checks (join verifiers, aggregation/constraint audits) for precision-limited systems.

\textit{\texttt{Pass@k} does not fix low-recall inputs.}
Increasing k changes candidate breadth, but the schema recall per outcome remains essentially flat across k (e.g., \textsc{C3-SQL} correct recall 0.985→0.982 from k=1 to 20; wrong recall 0.929→0.927). Thus, more samples cannot compensate for a missing schema. If low recall is detected, the system should revisit the schema step (e.g., raise the table/column budget, rerun selection with different cues) before spending tokens on additional CG/QR attempts.

\begin{table}[H]
  \centering
  \caption{Schema selection recall (higher is better) under four scenarios: Candidate Generation module generates correct SQL, wrong SQL, and Query Revision module generates correct SQL, wrong SQL. For methods producing multiple candidates, we list results at different \texttt{Pass@k}.}
  \label{tab:recall_bird_deepseekv3}
  \setlength\tabcolsep{2pt}
  \renewcommand{\arraystretch}{0.96}
  \scriptsize
  \resizebox{\linewidth}{!}{%
    \begin{tabular}{l|cc|cc}
      \toprule
      \multirow{2}{*}{\makecell{Approach}} &
        \multicolumn{2}{c|}{Candidate Generation} &
        \multicolumn{2}{c}{Query Revision} \\
      \cmidrule(lr){2-3}\cmidrule(lr){4-5}
        & \makecell{Correct\\Recall}
        & \makecell{Wrong\\Recall}
        & \makecell{Correct\\Recall}
        & \makecell{Wrong\\Recall} \\
      \midrule
      C3\mbox{-}SQL (Pass@1)          & 0.9848 & 0.9291 & 0.9837 & 0.9300 \\
      C3\mbox{-}SQL (Pass@5)          & 0.9834 & 0.9277 & 0.9837 & 0.9300 \\
      C3\mbox{-}SQL (Pass@10)         & 0.9837 & 0.9268 & 0.9837 & 0.9300 \\
      C3\mbox{-}SQL (Pass@15)         & 0.9829 & 0.9269 & 0.9837 & 0.9300 \\
      C3\mbox{-}SQL (Pass@20)         & 0.9817 & 0.9273 & 0.9837 & 0.9300 \\
      DIN\mbox{-}SQL                  & 0.9332 & 0.8203 & 0.9316 & 0.8215 \\
      MAC\mbox{-}SQL                  & 0.9946 & 0.9763 & 0.9930 & 0.9778 \\
      CHESS\textsubscript{(IR,SS,CG)} & 0.9197 & 0.8209 & 0.9135 & 0.8218 \\
      TA\mbox{-}SQL                   & 0.9466 & 0.8500 & --     & --     \\
      GSR                             & 0.9208 & 0.8142 & 0.9142 & 0.8132 \\
      RSL\mbox{-}SQL                  & 0.9494 & 0.9204 & 0.9495 & 0.8986 \\
      OpenSearch\mbox{-}SQL (Pass@1)  & 0.9831 & 0.9547 & 0.9811 & 0.9572 \\
      OpenSearch\mbox{-}SQL (Pass@5)  & 0.9821 & 0.9538 & 0.9811 & 0.9572 \\
      OpenSearch\mbox{-}SQL (Pass@10) & 0.9815 & 0.9543 & 0.9811 & 0.9572 \\
      OpenSearch\mbox{-}SQL (Pass@15) & 0.9814 & 0.9540 & 0.9811 & 0.9572 \\
      OpenSearch\mbox{-}SQL (Pass@20) & 0.9812 & 0.9538 & 0.9811 & 0.9572 \\
      \bottomrule
    \end{tabular}%
  }
\end{table}

\textit{\textbf{Insight: }Schema recall is a important indicator of end-to-end success. A shortfall at the schema selection stage propagates and is rarely corrected later; pipelines should monitor recall and re-enter schema selection when coverage is low. Additionally, More candidates are not a remedy for a missing schema.}